\documentclass[10pt,twocolumn,english,letterpaper,aps,prb,superscriptaddress]{revtex4}
\usepackage[T1]{fontenc}
\usepackage[latin9]{inputenc}
\usepackage{amsmath}
\usepackage{amssymb}
\usepackage{yfonts}
\usepackage{epsfig}

\newcommand{\usewidetext}[1]{}
\newcommand{\nowidetext}[1]{#1}

\newcommand{\secref}[1]{Sec.~\ref{#1}}
\newcommand{\Secref}[1]{Section \ref{#1}}
\newcommand{\secsand}[2]{Sections \ref{#1} and \ref{#2}}

\newcommand{\apref}[1]{Appendix \ref{#1}}
\newcommand{\apsand}[2]{Appendices \ref{#1} and \ref{#2}}
\newcommand{\Eqref}[1]{Equation~(\ref{#1})}
\renewcommand{\eqref}[1]{Eq.~(\ref{#1})}
\newcommand{\eqsref}[1]{Eqs.~(\ref{#1})}
\newcommand{\Eqsand}[2]{Equations~(\ref{#1}) and (\ref{#2})}
\newcommand{\eqsand}[2]{Eqs.~(\ref{#1}) and (\ref{#2})}
\newcommand{\Eqsdash}[2]{Equations~(\ref{#1}--\ref{#2})}
\newcommand{\eqsdash}[2]{Eqs.~(\ref{#1}--\ref{#2})}
\newcommand{\exref}[1]{(\ref{#1})}
\newcommand{\exsdash}[2]{(\ref{#1}--\ref{#2})}

\newcommand{\bea}{\begin{eqnarray}}
\newcommand{\eea}{\end{eqnarray}}
\newcommand{\beq}{\begin{equation}}
\newcommand{\eeq}{\end{equation}}
\newcommand{\lt}{\left}
\newcommand{\rt}{\right}
\newcommand{\la}{\lt\langle}
\newcommand{\ra}{\rt\rangle}
\newcommand{\dd}{\partial}
\newcommand{\const}{{\rm const}}
\newcommand{\eps}{\epsilon}

\newcommand{\kpar}{k_\parallel}
\newcommand{\kperp}{k_\perp}
\newcommand{\vths}{v_{\mathrm{th}s}}
\newcommand{\vthi}{v_{\mathrm{th}i}}
\newcommand{\vthe}{v_{\mathrm{th}e}}

\newcommand{\mfpi}{\lambda_{\mathrm{mfp}i}}
\newcommand{\mfpe}{\lambda_{\mathrm{mfp}e}}
\newcommand{\dBperp}{\delta B_\perp}
\newcommand{\dBpar}{\delta B_\parallel}
\newcommand{\dTe}{\delta T_{\parallel e}}
\newcommand{\dpe}{\delta p_{\parallel e}}
\newcommand{\Apar}{A_\parallel}
\newcommand{\Apari}{A_{\parallel i}}
\newcommand{\Apare}{A_{\parallel e}}
\newcommand{\Epar}{E_\parallel^{\rm st}}
\newcommand{\jpar}{j_\parallel}
\newcommand{\uperp}{u_\perp}
\newcommand{\upare}{u_{\parallel e}}
\newcommand{\upari}{u_{\parallel i}}
\newcommand{\vpar}{v_\parallel}
\newcommand{\hg}{\hat{g}}
\newcommand{\tg}{\tilde{g}}
\newcommand{\hh}{\tilde{h}}

\newcommand{\hvpar}{\hat{v}_\parallel}
\newcommand{\vperp}{v_\perp}
\newcommand{\vx}{\mathbf{r}}
\newcommand{\vv}{\mathbf{v}}
\newcommand{\vRs}{\mathbf{R}_s}
\newcommand{\vRe}{\mathbf{R}_e}
\newcommand{\vRi}{\mathbf{R}_i}
\newcommand{\vdel}{\mathbf{\nabla}}
\newcommand{\vb}{\hat{\mathbf{b}}}
\newcommand{\vk}{\mathbf{k}}
\newcommand{\rhos}{\rho_s}
\newcommand{\kape}{\kappa_{\parallel e}}
\newcommand{\vu}{\mathbf{u}}
\newcommand{\tphi}{\tilde\varphi}
\newcommand{\tchi}{\tilde\chi}
\newcommand{\tlambda}{\tilde\lambda}
\newcommand{\vueff}{\vu_{\mathrm{eff}}}
\newcommand{\vB}{\mathbf{B}}
\newcommand{\vy}{\hat{\mathbf{y}}}
\newcommand{\vz}{\hat{\mathbf{z}}}
\newcommand{\dvB}{\delta\vB_\perp}
\newcommand{\dB}{\delta B}
\newcommand{\IG}{I_G}

\newcommand{\Ei}{K_i}
\newcommand{\Ee}{K_e}
\newcommand{\Enst}{Y}
\newcommand{\Astuff}{X}
\newcommand{\We}{H_e}
\newcommand{\Wf}{W_{\rm fluid}}
\newcommand{\Um}{U}

\newcommand{\Dcoll}{D_{\rm coll}}
\newcommand{\flux}{\Gamma}
\newcommand{\adex}{G}
\newcommand{\mc}{m_c}
\newcommand{\mg}{m_\gamma}
\newcommand{\din}{\delta_\mathrm{in}}
\newcommand{\dres}{\delta_\eta}

\makeatletter

\usepackage{geometry}

\@ifundefined{definecolor}
 {\usepackage{color}}{}

\makeatother

\usepackage{babel}

\begin{document}

\title{Reduced fluid-kinetic equations for low-frequency dynamics, 
magnetic reconnection and electron heating in low-beta plasmas}
\author{Alessandro Zocco}
\email{a.zocco1@physics.ox.ac.uk}
\affiliation{Euratom/CCFE Fusion Association, 
Culham Science Centre, Abingdon OX14 3DB, UK}
\affiliation{R.~Peierls Centre for Theoretical Physics, 
University of Oxford, Oxford OX1 3NP, UK}
\author{Alexander A.\ Schekochihin}
\email{a.schekochihin1@physics.ox.ac.uk}
\affiliation{R.~Peierls Centre for Theoretical Physics, 
University of Oxford, Oxford OX1 3NP, UK}
\date{19 October 2011 published in {\em Physics of Plasmas}~{\bf 18}, 102309 (2011)}

\begin{abstract}
A minimal model for magnetic reconnection and, generally, low-frequency 
dynamics in low-beta plasmas is proposed. The model combines 
analytical and computational simplicity with physical realizability: 
it is a rigorous limit of gyrokinetics for plasma beta of order 
the electron-ion mass ratio. The model contains collisions and 
can be used both in the collisional and collisionless reconnection 
regimes. It includes gyrokinetic ions (not assumed cold) and allows 
for the topological rearrangement of the magnetic field lines by 
either resistivity or electron inertia, whichever predominates. 
The two-fluid dynamics are coupled to electron kinetics --- 
electrons are not assumed isothermal and are described by a reduced 
drift-kinetic equation. The model therefore 
allows for irreversibility and conversion of magnetic energy into 
electron heat via parallel phase mixing in velocity space. 
An analysis of the exchanges between various forms of free energy  
and its conversion into electron heat is provided. It is shown how
all relevant linear waves and regimes of the tearing instability 
(collisionless, semicollisional and fully resistive) are recovered 
in various limits of our model. 
An efficient way to simulate our equations numerically is proposed,
via the Hermite representation of the velocity space. 
It is shown that small scales in velocity space will form, giving 
rise to a shallow Hermite-space spectrum, whence it is inferred that, 
for steady-state or sufficiently slow dynamics, 
the electron heating rate will remain finite in the limit 
of vanishing collisionality. 
\end{abstract}

\pacs{52.30.Gz, 52.35.Vd}
\maketitle

\section{Introduction\label{sec:intro}}

Multiscale nonlinear plasma phenomena pose 
some of the most tantalizing and intellectually challenging 
theoretical problems in the field. Roughly speaking, this is because 
they tend to involve coupling between fluid (large) and kinetic (small) 
scales and so strain both our physical intuition (which in most cases is 
anchored in the fluid description) and analytical stamina (which tends 
to break down when full Vlasov-Maxwell kinetic theory appears to be 
the only rigorous recourse). 
A particularly clear example of this situation is the development
of the (nonlinear) kinetic theory of magnetic reconnection.

Magnetic reconnection, the unfreezing of magnetic flux in nonideal conducting 
plasmas,\cite{bisk-book-00,yamada_review} is one of those fundamental plasma physical processes 
understanding which turns out to be an essential theoretical building block in 
a vast range of laboratory, space and astrophysical applications: the sawtooth crash 
in tokamaks,\cite{Jim} solar flares,\cite{giovannelli,sweet_review} 
the magnetosheath and magnetotail of the Earth,\cite{dungey,xiao,eastwood:6235}
various types of magnetohydrodynamic and plasma turbulence,\cite{retino,servidio:115003} 
particle acceleration mechanisms,\cite{drake_nature,drake:963} 
as well as a plethora of more exotic reconnection-related phenomena believed to occur 
in extreme astrophysical environments\cite{uzdensky-review} 
--- solutions to these problems turn out to depend crucially 
on whether one thinks magnetic reconnection is fast or slow, 
steady or bursty and impulsive,\cite{bhatta_review} 
collisional or collisionless,\cite{cassak05,uzdensky:2139} 
essentially two- or three-dimensional, mediated by resistivity,\cite{sweet,parker} 
dispersive waves,\cite{rogers-prl-01-disp-waves} 
instabilities of the current sheets,\cite{Loureiro_07,Samtaney_09,Bhatta_09,Cassak_09,Huang_10,uls}
turbulence,\cite{matthaeus:2513,kowal-2009,lapenta:235001,Loureiro_09} or various combinations 
of these. 

There appears to be a broad theoretical consensus that in many 
natural plasmas, reconnection cannot be fully understood in a purely 
magnetohydrodynamic (MHD) setting (the same is true for most other 
nonlinear multiscale plasma phenomena). 
In recent years, a dramatic increase in the raw computing power that 
could be brought to bear on these problems has fuelled an intense 
effort --- and some spectacular empirical progress --- in understanding kinetic 
reconnection using PIC 
simulations,\cite{gem,ricci:4102,daughton:072101,drake:042306,daughton:065004,daughton:072117} 
both in two and, more recently, three dimensions.
One limitation of this approach is the lack of a simple analytical
framework that can help in the interpretation of numerical results.
Ever since the pure MHD approach was recognized as insufficient, 
and even before that, attempts have continued to develop various minimal 
models, usually of the two-fluid kind,\cite{hazeltine:3204,schep-phpl-94-twofluid,fitzpatrick:4713,fitzpatrick-err,waelbroeck:032109,tassi:034007} 
with the aim of capturing the ``essential physical ingredients,'' reflecting 
at any given time the evolving understanding of what those were. 
Finding such minimal models has been and is likely to remain
necessary also because, despite the rapid expansion of brute-force 
simulations, the resolution available, however impressive, is, in fact, 
never enough: the problem has at least three well separated spatial 
scales (global fluid, ion kinetic and electron kinetic), as well 
as possibly a similar level of complexity in the kinetic phase 
(velocity) space, and is likely to require three dimensions in an 
essential way. Furthermore, the current fully kinetic PIC simulations\cite{ricci:4102} 
have as yet to produce a clear picture of reconnection in low-beta 
plasmas embedded in a strong guide field, while gyrokinetic 
simulations of magnetic reconnection\cite{wan:012311,rogers:092110,wang:072103,perona:042104,numata} 
are in their infancy and are afflicted both by high numerical 
resolution requirements and lack of clarity on the theoretically 
expected outcomes (the gyrokinetic theory\cite{frieman:502} 
in its general form is not much less analytically difficult that the full Vlasov-Maxwell
kinetics, although numerically it is, of course, a radical simplification). 

There is a good reason to believe that a physically realizable model 
set of equations 
for weakly collisional reconnection, or nonlinear plasma dynamics 
generally, however simplified, cannot in general be fluid. 
Like most nonlinear phenomena, magnetic reconnection can and, indeed, 
should be thought about in terms of energy conversion, 
namely, conversion of magnetic energy associated with the reconnecting 
large-scale configuration into other forms of plasma energy and, ultimately, 
particle heat. It is when the energy is dissipated into heat that 
this conversion becomes irreversible. However, when collisions are weak, 
Ohmic (resistive) heating is not an important process, so most purely 
fluid models of collisionless reconnection turn out to be 
Hamiltonian:\cite{hazeltine:3204,schep-phpl-94-twofluid,waelbroeck:032109,tassi:034007} 
reconnection converts magnetic energy into other forms of ``fluid'' 
energy (e.g., kinetic energy of the mean electron flows),
the entropy of the system does not increase,  
everything is in principle reversible, and there is no heating. 
This is somewhat similar to inviscid dynamics in a neutral fluid. 
In fact, just like in neutral fluids, nonlinear dynamics in plasmas 
generically trigger formation of small scales, so even small dissipation 
coefficients become non-negligible because they multiply large gradients
(which is why dynamics in fluids with small viscosity are not everywhere 
inviscid, boundary layers form). 
In weakly collisional plasmas, the small-scale structure arises 
in phase space (i.e., both in position and particle velocities) 
via linear and nonlinear phase mixing processes (see \secref{sec:Hspectrum}). 
Once large velocity-space 
gradients are present, even weak collisions are sufficient to dissipate 
energy and cause heating. Formally, this can be understood by 
tracing the evolution and flows of free energy --- 
the quadratic invariant whose transfer to small 
scales and eventual thermalization is the central process in the 
non-equilibrium thermodynamics of kinetic plasmas 
(in the context of kinetic and gyrokinetic turbulence, this is explained in 
Refs.~\onlinecite{PPCFalex,alex}, 
where further references are provided; in application to magnetic reconnection, 
we expand on this topic in some detail in \secref{sec:energetics}). 
Thus, we believe there is a need for a minimal model of weakly 
collisional reconnection that is {\em not} fully conservative, 
because exact conservation properties impose constraints on the phase space 
of the system\cite{cafaro-prl-98-hamilt,grasso-prl-01-ph_mix} 
the breaking of which is not just non-negligible but is in fact likely 
to be physically essential in real plasmas.\footnote{It is probably also 
essential, for a similar reason, that the model be three-dimensional; 
restriction to exact two-dimensionality imposes many additional conservation 
laws (see \apref{app:2D}).} 
While numerical evidence is perhaps not fully conclusive on this subject, 
a few recent studies have, explicitly or implicitly, stressed 
the importance of electron heating in the nonlinear regime 
of kinetic reconnection.\cite{bowers:035002,daughton:072117,perona:042104} 

In this paper, we take the view that a minimal model that is as fluid-like 
as possible is clearly desirable, but {\em ad hoc} closure approximations are dangerous 
even if they appear to be physically motivated. We would like therefore 
to have a model that combines analytical and computational simplicity with physical 
realizability, i.e., constitutes a rigorously correct approximation 
of the kinetic system in some well 
defined physical limit. It turns out such a model can indeed be constructed 
and is a simple generalization of an existing 
two-fluid model,\cite{schep-phpl-94-twofluid} although it is not a fluid model 
in that it does retain kinetic electrons: the equations are the continuity 
equation, the ``gyrokinetic Poisson equation'' for the ions, 
the generalized Ohm's law and a version of the drift-kinetic equation 
for electrons; the latter is coupled to the fluid variables via 
the gradient of parallel electron temperature (energetically, via work done 
by the parallel electron pressure gradient).
The formal limit in which this model is derived is a combination of the gyrokinetic 
regime (strong guide field, frequencies below ion cyclotron, strong 
anisotropy $\kpar\ll\kperp$, small-amplitude fluctuations)
with a low-beta expansion (plasma beta of order electron-ion mass ratio)
--- this is explained in detail in \secref{sec:order}. 
A very close precursor of our approach in the existing literature 
is the fluid-kinetic model proposed by de Blank,\cite{blank:3927,blank:309} 
which couples a two-fluid model to a simplified 
electron kinetic equation, although he does not give a rigorous 
asymptotic ordering under which his model holds and also restricts 
his attention to the exactly 2D, exactly collisionless case, which 
means that his model is Hamiltonian and has an infinite 
number of Lagrangian conserved quantities\cite{liseikina:3535,pegoraro:243} 
(see \apref{app:2D}).

Let us discuss the basic physical ingredients that are retained 
in our approach. Despite the simplicity of our 
equations (summarized in \secref{sec:summary}), we believe 
that very little is lost of what we
consider the {\em sine qua non} of kinetic reconnection. 

\paragraph*{Three scales.} The problem is fundamentally three-scale: 
the model has to allow for a reconnecting configuration 
on fluid scales, dispersive effects at the ion scales and the flux unfreezing 
by a collisionless mechanism associated with electron microphysics 
and so residing at electron scales.  
In our case, the ion scales are the ion sound and Larmor radii
(the ions are gyrokinetic and not assumed cold; see \secref{sec:ions}). 
The flux unfreezing is effected by electron inertia --- see \secref{sec:flux}; 
we will sacrifice the electron-Larmor-radius effects. 

\paragraph*{Collisions.} It is desirable for a good model to contain 
a smooth transition from collisionless to semicollisional to 
fully collisional (resistive MHD) regime --- both because this provides 
a way to benchmark against situations that are better understood and because 
theoretically it is possible that many natural systems teeter at the 
boundary between the collisionless and collisional regimes.\cite{cassak05,uzdensky:2139}  
The transition between the two has been the focus of several 
recent studies, both experimental\cite{yamada06} and 
theoretical,\cite{daughton:065004,daughton:072117,zocco:110703,numata} 
but remains poorly understood. In our model, electron-ion collisions 
are retained (see \secref{sec:colls})
and so both Ohmic resistivity and (parallel) electron 
heat conduction (in the semicollisional limit; see \secref{sec:semicoll}) 
are recoverable (but not ion or electron viscosity, a limitation that will 
be discussed further in \secref{sec:concs}). 

\paragraph*{Free-energy flows, Landau damping and electron heating.} Finally, 
as we explained above, the model provides an electron heating channel, 
operative even with very weak collisions, in the form of the coupling 
of the generalized Ohm's law to the electron kinetic equation with a 
collision operator retained. This means 
that the electron Landau damping (whose essential byproduct is parallel 
phase mixing) is included and free energy can be 
converted from various fluid forms into the electron entropy and thence 
to heat (see \secref{sec:enex}). Note that neither ion Landau damping 
nor the nonlinear perpendicular phase mixing of either 
species\cite{PPCFalex,alex,tatsuno:015003} survive in our model.\\

The rest of the paper is organized as follows. 
In \secref{sec:prelims}, we review the gyrokinetic system, which 
is our starting point (\secref{sec:gk}), and introduce the set of 
ordering assumptions that encode the physical limit in which our 
equations hold rigorously (\secref{sec:order}). 
In \secref{sec:eqns}, we derive the equations themselves
(they are summarized in \secref{sec:summary}). 
In \secref{sec:energetics}, we work out the energetics 
of these equations: the various forms the free energy takes (\secref{sec:free}), 
how it is exchanged between fields (\secref{sec:enex}) 
and what that implies about irreversibility, 
thermalization (dissipation), and electron heating (\secref{sec:heating}). 
In \secref{sec:phmix}, we introduce a spectral representation 
of the electron kinetics in terms of Hermite polynomials (using a modified 
Lenard-Bernstein operator for electron collisions, introduced in 
\secref{sec:colls}). This 
provides what appears to be both a remarkably simple computational 
approach and an intuitively appealing physical interpretation 
of velocity-space dynamics and electron heating as a cascade 
in Hermite space (\secref{sec:Hspectrum}). We are led to the 
conclusion that the electron heating rate should remain finite 
in the limit of positive but vanishing electron collision frequency 
--- except for fast-growing solutions like the tearing 
mode (\secref{sec:heatingrate}). 
We also derive the semicollisional limit of our equations\cite{braginskii}  
in \secref{sec:semicoll}. 
\Secref{sec:concs} contains the concluding discussion. 
The paper is supplemented with two technical appendices: 
on two-dimensional invariants of our system (\secref{app:2D}) 
and on the linear theory 
(gyrokinetic electron-Landau-damped Alfv\'en waves,\cite{howes} 
collisional (resistive MHD),\cite{FKR,coppi-res,ABC,drake:1777} 
semicollisional\cite{drake:1341,cowley:3230,pegoraro:647,pegoraro:364,aydemir:3025} 
and collisionless\cite{laval-pellat-vuillemin,drake:1341,drake:1777,basu:465,cowley:3230,porcelli:425,zakharov:3285,kuvshinov:867,mirnov:4468} 
tearing modes). 

We have adopted a rather gradual, step-by-step approach to the derivation 
of all results. An impatient reader, or one already familiar with 
most of the (standard) analytical machinery deployed here, can gain a 
basic idea of the main results and conclusions 
by looking at \secref{sec:summary}, \secref{sec:heating},
\secref{sec:heatingrate} and \secref{sec:concs}.

\section{Preliminaries\label{sec:prelims}}

\subsection{Gyrokinetics in a slab\label{sec:gk}}

The starting point for our derivation is the gyrokinetic 
description of magnetized plasma,\cite{frieman:502} 
which is appropriate for low-frequency ($\omega\ll\Omega_s$) 
anisotropic ($\kpar\ll\kperp$) fluctuations in the presence 
of a mean magnetic field. The simplest case, which is all we will 
require here, is that of a static
uniform equilibrium with zero electric field  
and a constant straight magnetic field 
$\mathbf{B}_{0}=B_{0}\hat{\mathbf{z}}$, 
whose direction defines the $z$ axis.  
The subscripts $\perp$ and $\parallel$ will always refer to directions
with respect to this equilibrium (``guide'') magnetic field. 
The gyrokinetic equations are derived by performing an expansion of the Vlasov-Landau
equation in the small parameter $\eps=\kpar/\kperp$,
and averaging out the particle Larmor motion, i.e., 
all frequencies greater than the cyclotron frequency $\Omega_s=q_sB_0/m_s c$, 
where $s=i,e$ is the species index, $q_s$ is particle charge, 
$m_s$ particle mass, and $c$ the speed of light. 
For a detailed derivation of homogeneous gyrokinetics in a slab,
the reader may consult Ref.~\onlinecite{howes}. 

Let us summarize the resulting equations. 
The distribution function up to first order in $\epsilon$ is 
\beq
f_{s}\lt(\mathbf{r},\vv,t\rt)=F_{0s}
-\frac{q_{s}\varphi(\mathbf{r},t)}{T_{0s}}F_{0s}
+h_{s}(\vRs,\vperp,\vpar,t),
\label{eq:fs}
\eeq
where the zeroth-order distribution  
\beq
F_{0s}(v)=\frac{n_{0s}}{\lt(\pi\vths^{2}\rt)^{3/2}}
\exp\lt[-\frac{\vpar^{2}+\vperp^{2}}{\vths^{2}}\rt]
\eeq
is a Maxwellian with uniform density $n_{0s}$
and temperature $T_{0s}$, 
$\vths=(2T_{0s}/m_s)^{1/2}$ is the thermal speed, 
$q_{s}\varphi/T_{0s}=O(\epsilon)$
is the Boltzmann response containing the electrostatic potential $\varphi$, 
and $h_{s}$ is the gyrocenter distribution function, 
also $O(\epsilon)$.  
Spatially, $h_{s}$ is defined as a function of the 
the gyrocenter, or guiding-center, coordinate
\beq
\vRs=\mathbf{r}+\frac{\vv_{\perp}\times\hat{\mathbf{z}}}{\Omega_{s}}
\eeq
and satisfies the gyrokinetic equation
\begin{align}
\nonumber
\frac{\dd h_{s}}{\dd t}+\vpar\frac{\dd h_{s}}{\dd z}
+\frac{c}{B_{0}}\lt\{ \la\chi\ra _{\vRs},h_{s}\rt\} =\\
\frac{q_{s}F_{0s}}{T_{0s}}\frac{\dd\la \chi\ra _{\vRs}}{\dd t}
+\lt(\frac{\dd h_{s}}{\dd t}\rt)_{c},
\label{eq:g1}
\end{align}
where $\chi(\mathbf{r},\vv,t)=\varphi-\vv\cdot\mathbf{A}/c$
is the gyrokinetic potential (containing the information about the electromagnetic 
field in the form of the scalar potential $\varphi$ and vector potential $\mathbf{A}$
of the perturbed magnetic field, $\delta\mathbf{B}=\vdel\times\mathbf{A}$), 
\beq
\lt\{ \la \chi\ra _{\vRs},h_{s}\rt\} 
=\hat{\mathbf{z}}\cdot\lt(\frac{\dd\la \chi\ra _{\vRs}}{\dd\vRs}\times\frac{\dd h_{s}}{\dd\vRs}\rt)
\eeq
is the Poisson brackets, 
$(\dd h_{s}/\dd t)_{c}$ is the (gyroaveraged)
collision operator, and 
\beq
\la \chi(\mathbf{r},\vv,t)\ra _{\vRs}
=\frac{1}{2\pi}\int_{0}^{2\pi}d\vartheta\,
\chi\lt(t,\vRs-\frac{\vv_{\perp}\times\hat{\mathbf{z}}}{\Omega_{s}},\vv\rt)
\eeq
is the average of $\chi$ at constant $\vRs$ over the gyroangles $\vartheta$.
In Fourier space, this gyroaveraging operation takes a simple mathematical form 
in terms of Bessel functions $J_0$ and $J_1$, so the Fourier transform of 
$\la \chi(\mathbf{r},\vv,t)\ra _{\vRs}$ 
with respect to $\vRs$ can be written as follows
\begin{eqnarray}
\nonumber
\la \chi\ra _{\vRs,\vk} & = & 
J_{0}(a_{s})\lt(\varphi_{\vk}-\frac{\vpar A_{\parallel\vk}}{c}\rt)\\
 &  & +\,\,\frac{T_{0s}}{q_{s}}\frac{2\vperp^{2}}{\vths^{2}}\frac{J_{1}(a_{s})}{a_{s}}
\frac{\delta B_{\parallel\vk}}{B_{0}},
\label{eq:chi}
\end{eqnarray}
where $a_{s}=\kperp \vperp/\Omega_{s}$, and 
$\varphi_{\vk}$, $A_{\parallel\vk}$ 
and $\delta B_{\parallel\vk}$ are Fourier transforms 
(with respect to $\mathbf{r}$) of the scalar potential, 
parallel component of the vector potential and 
parallel component of the perturbed magnetic field, respectively. 
These fields are determined via Maxwell's equations, namely, 
the quasineutrality and Amp\`ere's law, where particle densities 
and currents are calculated from the gyrocenter distribution $h_{s}$. 
These equations will be introduced in \secsand{sec:els}{sec:ions}, where 
we will need them to compute electron flow velocity 
and density perturbation. 

\subsection{Low-beta ordering\label{sec:order}}

We would like to derive a minimal model suitable for an analytical 
description of magnetic reconnection in the presence of 
a mean field, i.e., reconnection of antiparallel perturbations 
$\delta\mathbf{B}_\perp = -\hat{\mathbf{z}}\times\vdel\Apar$. 
Since we wish to have a model that describes a real physical situation, 
we cannot resort to writing {\em ad hoc} fluid equations. 
Instead, our model will take the form of an asymptotic expansion 
of the gyrokinetic equations under an appropriate physically motivated 
ordering of all spatial and time scales and of the perturbation amplitudes. 
In devising our ordering, we are guided by what is known or expected 
about the physical effects that are essential in any description 
of a kinetic reconnection process. 

\subsubsection{Spatial scales\label{sec:spscales}}

Firstly, in the presence of a strong guide field, 
electron inertia is expected 
to be a key mechanism for breaking the magnetic field lines
(flux unfreezing) in collisionless or weakly 
collisional plasma.\cite{FKR,Vas,Wesson,bulanov-phfl-92-emhd,ottaviani-porcelli} 
Thus, we order 
\beq
\kperp d_{e}\sim1,
\label{eq:de}
\eeq
where $d_{e}=c/\omega_{pe}=\rho_{e}/\sqrt{\beta_{e}}$ is the electron 
inertial scale and, to fix standard notation, 
$\omega_{pe}=(4\pi n_{0e}e^{2}/m_{e})^{1/2}$ is the electron
plasma frequency, $e=|q_e|$ the elementary charge, 
$\rho_{e}=\vthe/\Omega_{e}$ the electron Larmor radius,
and $\beta_{e}=8\pi n_{0e}T_{0e}/B_{0}^{2}$ the electron beta. 

Secondly, a key feature of kinetic reconnection is a double layer 
resulting from the decoupling of the electrons from the ions at the ion 
sound scale.\cite{bisk-book-00} To retain this effect, we order 
\beq
\kperp\rhos\sim1,
\label{eq:rhos}
\eeq
where $\rhos = \rho_i\sqrt{Z/2\tau}$ is the ion sound radius 
(subscript $s$ for ``sound'' not to be confused with the species index!), 
$\rho_{i}=\vthi/\Omega_{i}$ is the ion Larmor radius, 
$Z=q_{i}/e$ and $\tau=T_{0i}/T_{0e}$. 
We will consider the temperature ratio to be 
order unity, so \eqref{eq:rhos} immediately implies 
\beq
\tau\sim 1,\qquad
\kperp\rho_i\sim\lt(\frac{\tau}{Z}\rt)^{1/2}\sim1.
\label{eq:tau}
\eeq
Thus, we retain the ion FLR along with the ion sound scale. 
However, a further simplifying option remains open: 
assuming cold ions, $\tau\ll1$, we can eliminate the 
ion FLR effects. It is with a view to this possibility that 
we will carry the $\tau$ dependence in all our orderings 
discussed below. 

In order for \eqsand{eq:de}{eq:rhos} 
to be consistent, we must have, within our ordering, $d_e\sim\rhos$
(this does not mean that these scales must be similar, just that 
we are not hard-wiring into our model a disparity between them). 
This implies
\beq
\frac{d_e}{\rhos}=
\sqrt{2Z}\lt(\frac{m_{e}}{m_{i}}\rt)^{1/2}
\frac{1}{\sqrt{\beta_e}}\sim1.
\eeq
To achieve this, we order 
\beq
\beta_{e}\sim\frac{Zm_{e}}{m_{i}}\ll1.
\label{eq:beta}
\eeq
Thus, we are restricting our consideration to low-beta 
plasmas and allowing both species to have finite temperature. 
The ordering \exref{eq:beta} is appropriate, for example, 
for the solar corona\cite{uzdensky:2139} 
and low-beta laboratory experiments such as the 
LArge Plasma Device at UCLA.\cite{gekelman:2875} 
In modern tokamaks, values of $\beta_e\sim 10^{-3}$ 
can occur in the edge pedestal region in the H-mode regime.\cite{saibene:969} 

Note that in the gyrokinetic approximation, $\beta_e$ is 
considered order unity with respect to the ordering of 
all quantities in powers of $\eps=\kpar/\kperp$; 
we will treat our low-beta expansion as subsidiary to the 
gyrokinetic $\eps$ expansion. 

Now, using \eqref{eq:de}, we conclude
\beq
\kperp\rho_{e}\sim\sqrt{\beta_e}\ll1,
\label{eq:rhoe}
\eeq
which will allow us to neglect the electron FLR effects and derive 
an ``almost fluid'' set of equations for the electrons.

\subsubsection{Time scales and perturbation amplitudes}

Far from the reconnecting region, the plasma mass flow is ordered
with the $\mathbf{E}\times\mathbf{B}$ drift velocity. 
Thus, we order the fundamental time scale on which 
we allow our fields to vary in such a way that 
the characteristic frequency is that associated with the 
$\mathbf{E}\times\mathbf{B}$ velocity $\uperp$:
\beq
\omega\sim \kperp \uperp\sim \kperp^{2}\frac{c\varphi}{B_{0}}.
\label{eq:ExB}
\eeq
Note that the gyrokinetic approximation requires 
$\omega\ll\Omega_{i,e}$. 

Since it is an essential feature of our model to 
take into account electron kinetics, we must allow 
the parallel streaming frequency of the electrons 
to be the same order as the rate at which our fields vary, 
namely
\beq
\omega\sim \kpar\vthe.
\label{eq:kparvth}
\eeq

The requirement that \eqsand{eq:ExB}{eq:kparvth} should be 
consistent with each other imposes an ordering on 
the size of the scalar potential:
\beq
\frac{e\varphi}{T_{0e}}\sim
\frac{\kpar}{\kperp}\frac{1}{\kperp\rho_e}
\sim \frac{\eps}{\sqrt{\beta_e}},
\label{eq:phi}
\eeq
where we used \eqref{eq:rhoe}. 
Note that the appearance of the gyrokinetic expansion parameter 
$\eps$ in the ordering of the perturbation amplitudes 
confirms that the use of the gyrokinetic approximation
is appropriate in the physical circumstances we are considering. 

We further require that the density perturbations 
are of the same order as the electrostatic perturbations 
given by \eqref{eq:phi}:
\beq
\frac{\delta n_e}{n_{0e}}\sim\frac{Z}{\tau}\frac{e\varphi}{T_{0e}}
\sim \frac{Z}{\tau}\frac{\eps}{\sqrt{\beta_e}}
\label{eq:orderdne}
\eeq
(the factor of $Z/\tau$ will emerge in \secref{sec:ions} and 
is kept for book-keeping purposes).
Physically this follows from the requirement that 
the physics associated with the ion sound scale 
is retained [\eqref{eq:rhos}]. 

\subsubsection{Alfv\'enic perturbations}

We will now order the magnetic perturbations. 
Let us observe that the ordering \exref{eq:beta} implies 
that the electron thermal speed is comparable to the Alfv\'en 
speed $v_A=B_0/\sqrt{4\pi m_in_{0i}}$: since 
$n_{0i}=n_{0e}/Z$ (quasineutrality), we have
\beq
\frac{\vthe}{v_A} 
= \lt(\frac{\beta_e}{Z}\frac{m_i}{m_e}\rt)^{1/2}\sim1. 
\eeq
Therefore,
\beq
\omega\sim \kpar v_A,
\eeq
i.e., Alfv\'en waves can propagate along the guide field 
with the same characteristic frequency as the electrons 
stream and plasma flows. Note that in view of 
\eqsand{eq:rhos}{eq:tau}, 
this ordering holds both for the magnetohydrodynamic 
Alfv\'en waves ($\omega=\kpar v_A$), 
and for the kinetic Alfv\'en waves 
($\omega\sim\kpar v_A\kperp\rhos$). 

Stipulating that Alfv\'enic perturbations should be 
accommodated by our ordering (which is equivalent to demanding 
that the Lorentz force is nonnegligible --- an essential 
ingredient in a reconnecting system), we deduce the ordering for the 
perpendicular perturbed magnetic field 
$\delta\mathbf{B}_\perp=-\hat{\mathbf{z}}\times\vdel_\perp\Apar$: 
\beq
\frac{\delta B_\perp}{B_{0}}\sim 
\frac{\uperp}{v_A}\sim
\eps\,\kperp\rho_i\lt(\frac{Z}{\tau}\rt)^{1/2}
\sim\eps,
\label{eq:dBperp}
\eeq
where we have used \eqsand{eq:phi}{eq:tau}. 

Note that, since $\eps\sim\kpar/\kperp$, this implies 
$\kperp{\delta B_\perp}\sim\kpar B_0$, i.e., the 
spatial variation of all quantities along the exact magnetic field 
contains comparable contributions from their variation along 
both the mean and perturbed fields --- a general property of 
gyrokinetic perturbations. This is just what is needed for 
reconnection problems with a guide field because Alfv\'enic 
dynamics with respect to the perturbed field must be allowed. 

Finally, from the perpendicular component of 
Amp\`ere's law (see, e.g., Eq.~120 of Ref.~\onlinecite{alex}), 
\beq
\frac{\delta B_\parallel}{B_0}\sim \beta_e\,\frac{e\varphi}{T_{0e}}
\sim \eps\sqrt{\beta_e},
\label{eq:dBpar}
\eeq
where \eqref{eq:phi} has been used. 
This will cause the parallel perturbations of the magnetic field 
(i.e., perturbations of the field strength) to fall out of 
our final set of equations. Indeed, it is a well known fact that 
in the low-beta ordering such perturbations tend to be negligible. 

This completes the ordering of the perturbation amplitudes. 

\subsubsection{Resistivity and collisions}

While our main focus is on collisionless reconnection, we would 
like to make contact with the collisional limit, in which 
the magnetic flux unfreezing and magnetic energy release is 
accomplished by Ohmic resistivity.\cite{spitzer}
We can retain resistivity by ordering the electron-ion
(and consequently electron-electron) collision frequency as 
comparable to the characteristic frequency of all the other 
processes that we are taking into account:
\beq
\nu_{ei}=Z\nu_{ee}\sim\omega.
\label{eq:nue}
\eeq
Since the Ohmic magnetic diffusivity (often colloquially called resistivity) 
is $\eta\sim \nu_{ei}d_e^2$, the above ordering means that the 
diffusive effects are retained in our ordering: 
indeed, using \eqsand{eq:de}{eq:nue},
\beq
\eta\kperp^2\sim \nu_{ei}\kperp^2d_e^2 \sim \nu_{ei} \sim \omega. 
\eeq
The resistivity can later be neglected in a subsidiary 
collisionless expansion, but we consider it useful to have a model 
in which a smooth transition from collisional to collisionless 
regime is possible 
(cf.\ Refs.~\onlinecite{yamada06,daughton:065004,daughton:072117,numata} 
and ideas on the marginal collisional-collisionless reconnection 
regime\cite{cassak05,uzdensky:2139,zocco:110703}). 

Our ordering of the electron collision immediately implies an ordering 
of the ion collisions:\cite{Per}
\begin{eqnarray}
\label{eq:nuii}
\nu_{ii}&=& \frac{Z^2}{\tau^{3/2}}\lt(\frac{m_e}{m_i}\rt)^{1/2}\nu_{ei}
\sim \lt(\frac{Z}{\tau}\rt)^{3/2}\!\!\sqrt{\beta_e}\,\omega,\quad\\
\label{eq:nuie}
\nu_{ie}&=& \frac{Zm_e}{m_i}\,\nu_{ei} \sim \beta_e\,\omega,
\end{eqnarray}
where we have used \eqsand{eq:beta}{eq:nue}. 
Thus, we are effectively assuming ions to be collisionless (no ion viscosity). 

Finally, it is perhaps useful to note the ordering of the particle mean free path:
using \eqsand{eq:kparvth}{eq:nue}, we have
\beq
\kpar\mfpe = \lt(\frac{Z}{\tau}\rt)^2\kpar\mfpi = 
\frac{\kpar\vthe}{\nu_{ei}} \sim 1. 
\eeq

\section{Derivation of the equations\label{sec:eqns}}

We are now ready to apply our ordering and derive from the 
gyrokinetic system (\secref{sec:gk}) a reduced system of equations 
describing gyrokinetic ions and drift-kinetic electrons 
in a low-beta weakly collisional plasma. 
An impatient reader can skip to \secref{sec:summary}.

\subsection{Electrons\label{sec:els}}

In \eqref{eq:g1} for electrons ($s=e$), we retain only the 
lowest order in the expansion with respect to $\beta_e$. 
In order to do this, we note first that the Bessel functions 
in the expression for 
$\la \chi\ra _{\vRe}$
[\eqref{eq:chi}] can be expanded in small argument because 
$a_e\sim\kperp\rho_e\sim\sqrt{\beta_e}\ll1$ [\eqref{eq:rhoe}]: 
\beq
J_0(a_e)\simeq \frac{2J_1(a_e)}{a_e} = 1 + O(a_e^2). 
\eeq 
Using \eqsand{eq:phi}{eq:dBperp}, we find that the $\varphi$ 
and $\Apar$ terms in \eqref{eq:chi} are the same order,
\beq
\frac{\vpar\Apar}{c} \sim \frac{\vthe B_0}{c\kperp}\frac{\dBperp}{B_0}
\sim \frac{T_{0e}}{e}\frac{\eps}{\kperp\rho_e}\sim\varphi,
\label{eq:Aparels}
\eeq
while, according to \eqref{eq:dBpar}, the $\dBpar$ term 
is one order of $\beta_e$ smaller. Thus, 
\beq
\la \chi\ra _{\vRe} 
= \lt(\varphi - \frac{\vpar\Apar}{c}\rt) 
\lt[1+O(\beta_e)\rt].
\label{eq:chie}
\eeq

This allows us to write \eqref{eq:g1} as follows, up to 
corrections of order $O(\beta_e)$, 
\beq
\label{eq:egk}
\frac{dh_{e}}{dt}+\vpar\vb\cdot\vdel h_{e}
= -\frac{eF_{0e}}{T_{0e}}\frac{\dd}{\dd t}
\lt(\varphi-\frac{\vpar \Apar}{c}\rt)
+ \lt(\frac{\dd h_e}{\dd t}\rt)_{c},
\eeq 
where we have introduced a ``convective'' time derivative 
incorporating the $\mathbf{E}\times\mathbf{B}$ motion and 
the parallel spatial derivative along the perturbed field line:
\bea
\frac{dh_{e}}{dt} &=& 
\frac{\dd h_{e}}{\dd t} + \frac{c}{B_0}\,\bigl\{\varphi,h_e\bigr\},\\
\vb\cdot\vdel h_{e} &=& 
\frac{\dd h_e}{\dd z} - \frac{1}{B_0}\,\bigl\{\Apar,h_e\bigr\}.
\eea
Note that all terms in \eqref{eq:egk} are comparable under ordering 
because of the assumptions we made about the time scales for 
the $\mathbf{E}\times\mathbf{B}$ flows [\eqref{eq:ExB}], 
electron streaming [\eqref{eq:kparvth}], 
collisions [\eqref{eq:nue}] and perturbation amplitudes 
[\eqsand{eq:phi}{eq:dBperp}]. 

It is now convenient formally 
to split the electron distribution function in such a way 
as to separate the inhomogeneous solution arising from the first 
term on the right-hand side of \eqref{eq:egk} as well as the density 
and parallel velocity moments: 
\beq
h_{e}=\lt(-\frac{e\varphi}{T_{0e}}
+\frac{\delta n_{e}}{n_{0e}}
+\frac{2\vpar u_{\parallel e}}{\vthe^{2}}\rt)F_{0e}+g_{e},
\label{eq:eansatz}
\eeq
where by definition 
\bea
\frac{\delta n_{e}}{n_{0e}} &=& \frac{1}{n_{0e}}\int d^3\vv\,\delta f_e,\\
u_{\parallel e} &=& \frac{1}{n_{0e}}\int d^3\vv\,\vpar\delta f_e,
\eea
and $\delta f_e = h_e + e\varphi F_{0e}/T_{0e}$ is the 
total perturbed electron distribution function 
[see \eqref{eq:fs}]. 
Since the first term in \eqref{eq:eansatz} cancels the Boltzmann part in 
$\delta f_e$, the above definitions of the density and parallel electron 
flow velocity are consistent provided we demand that  
\beq
\int d^{3}\vv\lt(\begin{array}{c}1\\\vpar\end{array}\rt)
g_{e}=0,
\label{eq:properties}
\eeq
i.e., the ``reduced'' electron distribution function 
$g_e$ contains all higher moments of the electron distribution function, 
but not density or parallel flow.  

The perturbed density will be calculated in \secref{sec:ions} from the 
quasineutrality condition. The parallel electron velocity $\upare$ is related to 
$\Apar$ and to the parallel ion velocity $\upari$ via the parallel component 
of Amp\`ere's law:
\bea
\nonumber
\hat{\mathbf{z}}\cdot\lt(\vdel_\perp\times\delta\mathbf{B}_\perp\rt)
&=& - \nabla_\perp^2\Apar = \frac{4\pi}{c}\,\jpar\\ 
&=& \frac{4\pi e n_{0e}}{c}\lt(\upari-\upare\rt).
\label{eq:Ampere}
\eea
This can be easily manipulated into 
\beq
\upare = \frac{e}{cm_e}\,d_e^2\nabla_\perp^2\Apar + \upari.
\label{eq:upare}
\eeq

We now substitute \eqref{eq:eansatz} into \eqref{eq:egk} and 
take moments of the resulting equation. The zeroth moment is, 
after using \eqref{eq:upare}, 
\beq
\frac{d}{dt}\frac{\delta n_{e}}{n_{0e}}
+\vb\cdot\vdel\upari = 
- \vb\cdot\vdel 
\frac{e}{cm_e}\,d_e^2\nabla_\perp^2\Apar.
\label{eq:dne}
\eeq
The first $\vpar$ moment of \eqref{eq:egk} is, again using 
\eqref{eq:upare} and dividing through by $en_{0e}/cm_e$, 
\begin{align}
\nonumber
&\frac{d}{dt}\lt(\Apar-d_{e}^{2}\nabla_{\perp}^{2}\Apar\rt)=\\
\nonumber
&-c\,\frac{\dd\varphi}{\dd z} 
+ \frac{cT_{0e}}{e}\,\vb\cdot\vdel
\lt(\frac{\delta n_{e}}{n_{0e}}+\frac{\dTe}{T_{0e}}\rt)\\
&-\frac{cm_{e}}{en_{0e}}\int d^{3}\vv\,\vpar
\lt(\frac{\dd h_{e}}{\dd t}\rt)_{c} 
+ \frac{cm_e}{e}\frac{d\upari}{dt},
\label{eq:Apar}
\end{align}
where the parallel electron temperature perturbation has been 
introduced as a shorthand for the $\vpar^2$ moment of $g_e$: 
\beq
\frac{\dTe}{T_{0e}}=\frac{1}{n_{0e}}\int d^{3}\vv\,\frac{2\vpar^{2}}{\vthe^{2}}\,g_{e}.
\label{eq:dTe}
\eeq

Finally, an equation for $g_e$ is obtained from \eqref{eq:egk} 
by subtracting from it \eqref{eq:dne} multiplied by $F_{0e}$
and substituting $\dd\Apar/\dd t$ calculated using \eqref{eq:Apar}.
After a few lines of straightforward algebra, this gives 
\begin{align}
\nonumber
& \frac{dg_{e}}{dt}
+\vpar\vb\cdot\vdel
\lt(g_{e}- \frac{\dTe}{T_{0e}}\,F_{0e}\rt) - C[g_e] = \\
&\lt(1-\frac{2\vpar^{2}}{\vthe^{2}}\rt)F_{0e}
\vb\cdot\vdel
\lt(\frac{e}{cm_e}\,d_e^2\nabla_{\perp}^{2}\Apar + \upari\rt).
\label{eq:ge}
\end{align}
where the collisional terms have been assembled together: by 
definition,
\beq
C[g_e] = \lt(\frac{\dd h_{e}}{\dd t}\rt)_{c}
-\frac{2\vpar F_{0e}}{\vthe^{2}n_{0e}}
\int d^{3}\vv'\,v'_{\parallel}\lt(\frac{\dd h_{e}}{\dd t}\rt)_{c}.
\label{eq:Cdef}
\eeq
It is easy to verify that \eqref{eq:ge} respects the constraint 
on $g_e$ given by \eqref{eq:properties}. 
We stress that \eqref{eq:ge} is not homogeneous in $g_e$, so $g_e$ cannot be 
consistently neglected
(cf.\ Ref.~\onlinecite{schep-phpl-94-twofluid}). 

\subsection{Flux conservation\label{sec:flux}}

\Eqref{eq:Apar} can easily be rearranged into the following form
\beq
\frac{\dd\Apar}{\dd t} = -c\vb\cdot\vdel\tphi 
+ \eta\nabla_\perp^2\Apar + \frac{d}{dt}\,d_{e}^{2}\nabla_{\perp}^{2}\Apar,
\label{eq:Apar_flux}
\eeq
where 
\beq
\tphi \equiv \varphi 
- \frac{T_{0e}}{e}\lt(\frac{\delta n_{e}}{n_{0e}} + \frac{\dTe}{T_{0e}}\rt).
\eeq
Recalling that $\dvB=-\vz\times\vdel_\perp\Apar$ and $\vB = B_0\vz + \dvB$, 
we immediately infer from \eqref{eq:Apar_flux} that 
\beq
\frac{\dd\vB}{\dd t} = \vdel\times\lt(\vueff\times\vB\rt) + \eta\nabla_\perp^2\vB 
- \vz\times\vdel_\perp\frac{d}{dt}\,d_{e}^{2}\nabla_{\perp}^{2}\Apar,
\eeq
where $\vueff = \vz\times\vdel_\perp c\tphi/B_0$ is the effective velocity, 
into which, as the above equation demonstrates, the field lines are frozen
except for the resistive and electron-inertia effects. This is an application 
to our equations of the more general flux-conservation argument due to 
Cowley.\cite{cowley-thesis}

An important conclusion is that the kinetic electron effects, which enter via 
$\dTe$ determined from \eqsand{eq:dTe}{eq:ge}, do not unfreeze flux. 

\subsection{Ions\label{sec:ions}}

\Eqsdash{eq:dne}{eq:ge} are four equations that involve six 
quantities $\delta n_{e}$, $\varphi$, $\Apar$, $\upari$, $\dTe$
and $g_{e}$. In order to have a closed system, we need find 
additional equations for two of these quantities. The ion gyrokinetics 
will give us this additional information: namely, we will determine 
the ion parallel flow velocity $\upari$, which turns out to vanish to 
lowest order, and $\delta n_e/n_{0e}$, which, because of the quasineutrality, 
is the same as the ion density perturbation $\delta n_i/n_{0i}$. 

As we explained in \secref{sec:order}, the ions remain fully gyrokinetic under 
our ordering, i.e., the Bessel functions in the expression 
for $\la\chi\ra _{\vRi}$
[\eqref{eq:chi}] cannot be expanded in small argument because 
$a_i\sim\kperp\rho_i$ is finite [\eqref{eq:tau}]. However, a significant 
simplification of \eqref{eq:chi} for ions 
is possible because, as in the case of the electrons, 
the $\dBpar$ term is an order of $\beta_e$ smaller than the $\varphi$ 
term [\eqref{eq:dBpar}] and also because, unlike for the electrons, 
the $\Apar$ is also small: indeed, noticing that 
$\vthi/\vthe = (\tau m_e/m_i)^{1/2}$ and comparing with \eqref{eq:Aparels}, 
we find 
\beq
\frac{\vpar\Apar}{c} \sim \frac{\vthi B_0}{c\kperp}\frac{\dBperp}{B_0}
\sim \lt(\tau\,\frac{m_e}{m_i}\rt)^{1/2}\varphi
\sim \frac{\tau}{Z}\,\sqrt{\beta_e}\,\varphi.
\eeq
Thus, we have for the ions $\chi=\varphi$, or 
\beq
\la\chi\ra _{\vRi,\vk} = 
J_0(a_i)\varphi_{\vk}.
\eeq

Because of the way we ordered the electron streaming frequency  
[\eqref{eq:kparvth}] and the electron collisions [\eqref{eq:nue}], 
the ion streaming frequency 
\beq
\kpar\vthi = \lt(\tau\,\frac{m_e}{m_i}\rt)^{1/2} \kpar\vthe
\sim \frac{\tau}{Z}\,\sqrt{\beta_e}\,\omega 
\eeq
and the ion collisions 
[\eqsand{eq:nuii}{eq:nuie}] are small and so the corresponding 
terms in \eqref{eq:g1} for ions ($s=i$) are negligible to lowest order. 
What remains is the following rather simple equation for the ions, 
devoid of kinetic effects except for the ion FLR: 
\beq
\frac{\dd g_{i}}{\dd t} 
+ \frac{c}{B_{0}}\lt\{\la\varphi\ra_{\vRi},g_{i}\rt\} = 0,
\label{eq:gi}
\eeq
where we have introduced a new function
\beq
g_{i}=h_i - \frac{ZeF_{0i}}{T_{0i}}\la \varphi\ra_{\vRi}.
\label{eq:iansatz}
\eeq

\Eqref{eq:gi} is homogeneous and has one very straightforward solution: 
\beq
g_{i}=0.
\eeq
Thus, the ion response under our ordering is essentially electrostatic and 
the perturbed ion distribution function is [see \eqref{eq:fs}]
\beq
\delta f_i = \frac{Ze F_{0i}}{T_{0i}}
\lt(\la \varphi\ra_{\vRi} - \varphi\rt).
\label{eq:dfi}
\eeq
Note that the first term here is a function of the gyrocenter variable 
$\vRi$ while the second is a function of the position variable 
$\mathbf{r}$. When computing the ion density and flow velocity, we must 
integrate $\delta f_i$ over velocities while keeping $\mathbf{r}$ constant, 
which means that terms dependent on $\vRi$ must be gyroaveraged 
at constant $\mathbf{r}$. Just like the gyroaveraging at constant $\vRi$, 
this can be expressed in Fourier space as a multiplication by the Bessel 
function $J_0(a_i)$. Thus, we calculate
\bea
\nonumber
\frac{\delta n_i}{n_{0i}} &=& \frac{1}{n_{0i}}\int d^3\vv\,
\la\delta f_i\ra_{\mathbf{r}}\\
\nonumber
&=& \frac{Ze}{T_{0i}}\lt(\frac{1}{n_{0i}}\int d^3\vv\,
\la \la \varphi\ra_{\vRi} \ra_{\mathbf{r}} F_{0i}
- \varphi\rt)\\
&=& - (1-\hat{\Gamma}_0)\,\frac{Ze\varphi}{T_{0i}},
\label{eq:dni}
\eea
where $\hat{\Gamma}_0$ is the operator that is the inverse Fourier transform of
\beq
\Gamma_0(\alpha_i) = \frac{1}{n_{0i}}\int d^3\vv\, \lt[J_0(a_i)\rt]^2 F_{0i}
= I_0(\alpha_i) e^{-\alpha_i},
\label{eq:Gamma}
\eeq 
where $\alpha_i=\kperp^2\rho_i^2/2$ and $I_0$ is the modified Bessel function. 
By quasineutrality, \eqref{eq:dni} also gives us the electron density 
perturbation: 
\beq
\frac{\delta n_e}{n_{0e}} = - \frac{Z}{\tau}\,(1-\hat{\Gamma}_0)\frac{e\varphi}{T_{0e}}. 
\label{eq:Poisson}
\eeq
We have recovered a rather popular standard treatment of the ion FLR.\cite{krommes:1}
\Eqref{eq:Poisson} is often referred to as the gyrokinetic Poisson equation.
Note that \eqref{eq:Poisson} is consistent with the ordering 
for the density perturbation assumed in \secref{sec:order} [see \eqref{eq:orderdne}]. 

It will be useful to remember that for $\kperp\rho_i\ll1$, 
$\Gamma_0(\alpha_i)\simeq 1 - \alpha_i$ and so 
\beq
\frac{Z}{\tau}\,(1-{\Gamma}_0) \simeq
\kperp^2\rhos^2
\label{eq:Gamma_lw}
\eeq
in the long-wavelength limit --- this leads to a significant simplification 
of \eqref{eq:Poisson}.
The long-wavelength form of the gyrokinetic Poisson equation is useful when one 
wishes to neglect the ion FLR effects. Formally, this can be done by 
means of the cold-ion approximation, $\tau\ll1$. In view 
of \eqref{eq:tau}, it immediately implies $\kperp\rho_i\sim\sqrt{\tau}\ll1$.  
Note that taking this limit does not require us to order $\tau$ within 
the $\beta_e$ expansion as $\tau$ and $\beta_e$ do not interfere 
with each other in most of the ordering arguments of \secref{sec:order}. 
The only exception to this is \eqref{eq:nuii}, which implies that 
if $\tau$ is excessively small, ion collisions (and, therefore, ion viscosity) 
might become important. The condition for this {\em not} to happen is 
$\beta_e\ll\tau^3$, or $\tau\gg(m_e/m_i)^{1/3}$. 
Clearly, it is safest to keep $\tau$ only moderately small 
and to treat any possible expansion in $\tau$ as 
subsidiary to the $\beta_e$ expansion. 

Finally, since $\delta f_i$ given by \eqref{eq:dfi} is even in $\vpar$ and 
gyroaveraging only involves $\vv_\perp$, we have
\beq
\upari = \frac{1}{n_{0i}}\int d^3\vv\,\vpar
\la\delta f_i\ra_{\mathbf{r}} = 0. 
\label{eq:upari}
\eeq
This means that under the ordering we have adopted the parallel current 
is carried predominantly by the electrons [see \eqref{eq:Ampere}]. 

\Eqsand{eq:Poisson}{eq:upari} together with \eqsdash{eq:dne}{eq:ge} 
constitute a complete set. We will assemble and summarize them momentarily, 
but first let us work out the collisional terms in our equations. 

\subsection{Collisions\label{sec:colls}} 

In order to make our equations useable, we must calculate the collision terms. 
We would like to do this by employing a maximally simplified collision operator 
rather than the quantitatively correct but cumbersome Landau one. 
In order to choose the most appropriate model, we notice first that the 
electron kinetic equation does not involve any nontrivial evolution 
of the $\vperp$ structure in the distribution function. We therefore 
believe it is acceptable to treat the kinetic equation as one-dimensional 
in velocity space --- formally, this can be thought of as integrating out 
the $\vperp$ dependence in $g_e$ and $F_{0e}$. The simplest one-dimensional 
collision operator is the Lenard--Bernstein one,\cite{lenard:1456} which
we write in the following form:
\begin{align}
\nonumber
&\lt(\frac{\dd h_e}{\dd t}\rt)_c =\\ 
\nonumber
& \nu_{ei}\lt[\frac{1}{2}
\frac{\dd}{\dd\hvpar}\lt(\frac{\dd}{\dd\hvpar} + 2\hvpar\rt)h_e
+ \frac{2\hvpar\upari}{\vthe}\,F_{0e}\rt.\\
&\lt.+ \lt(1-2\hvpar^2\rt)F_{0e}\,\frac{1}{n_{0e}}
\int d\vpar'\,\lt(1-2\hvpar^{\prime 2}\rt)h_e\rt],
\end{align}
where $\hvpar=\vpar/\vthe$ and the additional, non-differential 
terms in the operator have been constructed in such a way that 
electron momentum evolution is captured correctly, particle number 
and parallel kinetic energy are conserved and the $H$ theorem 
is satisfied (cf.\ Ref.~\onlinecite{abel:122509}; 
the $H$ theorem will be proved explicitly in \secref{sec:Htheorem}). 
The collision frequency $\nu_{ei}$ is taken to be velocity-independent 
--- a gross simplification compared to the uncensored 
reality, in which it is in fact a strong function of velocity. 
We do not believe this to be an important shortcoming of our model, 
however, because our focus will be on the role of the collision operator 
as small-scale regularization in velocity space rather than on quantitatively 
precise calculations of the effects of finite collisionality.  

Let us now calculate the collisional terms in \eqsand{eq:Apar}{eq:ge}.
Using \eqref{eq:eansatz} for $h_e$ and \eqref{eq:Ampere} for $\upare$, 
we get
\bea
\nonumber
\frac{1}{n_{0e}}\int d^3\vv\,\vpar \lt(\frac{\dd h_e}{\dd t}\rt)_c 
&=& \nu_{ei}\lt(\upari-\upare\rt)\\
&=& - \frac{e\nu_{ei}}{cm_e}\, d_e^2\nabla_\perp^2\Apar,
\label{eq:res}
\eea 
which simply gives rise to a resistive diffusion term in \eqref{eq:Apar}
with Ohmic diffusivity $\eta = \nu_{ei}d_e^2$. 
The collision operator defined by \eqref{eq:Cdef} becomes
\bea
\nonumber
C[g_e] &=& 
\nu_{ei}\lt[\frac{1}{2}\frac{\dd}{\dd\hvpar}\lt(\frac{\dd}{\dd\hvpar} + 2\hvpar\rt)g_e
\rt.\\
&&\qquad - \lt. \lt(1-2\hvpar^2\rt)\frac{\dTe}{T_{0e}}\,F_{0e}\rt],
\label{eq:C_LB}
\eea 
where we have used \eqsand{eq:eansatz}{eq:res}, the properties of $g_e$ 
[\eqref{eq:properties}] and the definition of the parallel 
electron temperature perturbation [\eqref{eq:dTe}]. 

\subsection{Summary of the equations\label{sec:summary}}

Assembling now \eqsdash{eq:dne}{eq:ge}, \exref{eq:Poisson}, \exref {eq:upari} 
and \exref{eq:res}, we arrive at the following closed set 
\usewidetext{
\begin{widetext}
\begin{align}
\label{eq:phi_sum}
&\frac{d}{dt}\frac{Z}{\tau}\,(1-\hat{\Gamma}_0)\frac{e\varphi}{T_{0e}}=
\vb\cdot\vdel 
\frac{e}{cm_e}\,d_e^2\nabla_\perp^2\Apar,\\
\label{eq:Apar_sum}
&\frac{d}{dt}\lt(\Apar-d_{e}^{2}\nabla_{\perp}^{2}\Apar\rt)=
\eta\nabla_\perp^2\Apar - c\,\frac{\dd\varphi}{\dd z}
- \frac{cT_{0e}}{e}\,\vb\cdot\vdel
\lt[\frac{Z}{\tau}\,(1-\hat{\Gamma}_0)\frac{e\varphi}{T_{0e}}
-\frac{\dTe}{T_{0e}}\rt],\\
\label{eq:ge_sum}
&\frac{dg_{e}}{dt}
+\vpar\vb\cdot\vdel
\lt(g_{e}- \frac{\dTe}{T_{0e}}\,F_{0e}\rt) = C[g_e]
+ \lt(1-\frac{2\vpar^{2}}{\vthe^{2}}\rt)F_{0e}
\vb\cdot\vdel
\frac{e}{cm_e}\,d_e^2\nabla_{\perp}^{2}\Apar,
\end{align}
\end{widetext}
}
\nowidetext{
\begin{align}
\label{eq:phi_sum}
&\frac{d}{dt}\frac{Z}{\tau}\,(1-\hat{\Gamma}_0)\frac{e\varphi}{T_{0e}}=
\vb\cdot\vdel 
\frac{e}{cm_e}\,d_e^2\nabla_\perp^2\Apar,\\
\nonumber
&\frac{d}{dt}\lt(\Apar-d_{e}^{2}\nabla_{\perp}^{2}\Apar\rt)=
\eta\nabla_\perp^2\Apar - c\,\frac{\dd\varphi}{\dd z}\\ 
\label{eq:Apar_sum}
&- \frac{cT_{0e}}{e}\,\vb\cdot\vdel
\lt[\frac{Z}{\tau}\,(1-\hat{\Gamma}_0)\frac{e\varphi}{T_{0e}}
-\frac{\dTe}{T_{0e}}\rt],\\
\nonumber
& \frac{dg_{e}}{dt}
+\vpar\vb\cdot\vdel
\lt(g_{e}- \frac{\dTe}{T_{0e}}\,F_{0e}\rt) = C[g_e]\\
&+\lt(1-\frac{2\vpar^{2}}{\vthe^{2}}\rt)F_{0e}
\vb\cdot\vdel
\frac{e}{cm_e}\,d_e^2\nabla_{\perp}^{2}\Apar,
\label{eq:ge_sum}
\end{align}
}
where the following short-hand notation is used
\begin{align}
\label{eq:dTe_sum}
&\frac{\dTe}{T_{0e}}=\frac{1}{n_{0e}}\int d^{3}\vv\,\frac{2\vpar^{2}}{\vthe^{2}}\,g_{e},\\
&\frac{d}{dt} = 
\frac{\dd}{\dd t} + \frac{c}{B_0}\,\bigl\{\varphi,\dots\bigr\},\\
&\vb\cdot\vdel =
\frac{\dd}{\dd z} - \frac{1}{B_0}\,\bigl\{\Apar,\dots\bigr\},
\end{align}
$\eta=\nu_{ei}d_e^2$ is the Ohmic diffusivity 
and $C[g_e]$ is the collision operator --- a simple model for which, 
based on the Lenard--Bernstein operator, is given by \eqref{eq:C_LB}. 
These equations evolve three fields: $\varphi$, $\Apar$ and 
$g_e$, which are all functions of time and three spatial 
coordinates; $g_e$ is also a function of $\vpar$ and $\vperp$, 
although the $\vperp$ dependence can be ignored (or integrated 
out) if the Lenard--Bernstein collision operator is used. 

\Eqsdash{eq:phi_sum}{eq:ge_sum} constitute a minimal 
physically realizable paradigm for magnetic reconnection 
and, more generally, low-frequency nonlinear plasma dynamics with a strong guide field, 
including all effects we expect to be important: ion sound scale 
physics, ion FLR, electron inertia, electron collisions, Ohmic resistivity, 
and electron temperature perturbation determined by a kinetic equation. 
We will refer to these equations as 
{\em Kinetic Reduced Electron Heating Model (KREHM)}. 
Its hybrid fluid-kinetic nature and the presence of the kinetic electron 
heating channel (further discussed below), constitutes the main difference with 
previously considered models: in the 2D, collisionless case, our equations 
can be manipulated into a form similar to that proposed 
by de Blank\cite{blank:3927,blank:309}; by setting $g_e=0$ 
(isothermal-electrons closure), we recover 
the equations of Schep et al.\cite{schep-phpl-94-twofluid} 
While the use of a fluid model $(g_{e}=0)$ might be a useful simplification, 
we stress that setting $g_e=0$ cannot be rigorously justified (at least in 
the analytical framework we have chosen): indeed, $g_{e}=0$ is not a solution 
of \eqref{eq:ge_sum} unless $\vb\cdot\vdel\nabla_{\perp}^{2}\Apar=0$
(i.e., $\vb\cdot\vdel j_\parallel=0$), which cannot be the case in 
a reconnection-relevant solution. 

We will show in \secref{sec:energetics} 
that in the collisionless (or, more precisely, weakly collisional) 
limit, the coupling of the fluid system to the kinetic equation \exref{eq:ge_sum} 
via parallel electron temperature fluctuations
provides the electron heating channel and so makes collisionless
reconnection thermodynamically irreversible. We believe this to 
be fundamentally important and physically the most 
interesting outcome of our calculation. 
In the remainder of the paper  we will concentrate on this aspect.

We stress that although the kinetic coupling leads to irreversibility 
and heating, it does {\it not} constitute a flux-unfreezing mechanism by itself; 
magnetic field lines can only be broken by resistivity and electron inertia 
(a simple proof of this statement was provided in \secref{sec:flux}). 

To conclude this brief summary of our equations, let us note that 
a number of well-known limiting cases are easily derivable from them. 
As already discussed, in the collisionless limit, for $g_e=0$, 
they reduce to the two-fluid model for strong-guide-field collisionless 
reconnection.\cite{schep-phpl-94-twofluid} 
In the collisionally dominated limit, \eqref{eq:ge_sum} reduces to the 
standard equation for the evolution of the parallel electron temperature 
via parallel heat conduction (see \secref{sec:semicoll}). 
If collisions are so large that the resistive scale is larger than 
both $\rhos$ and $d_e$, \eqsand{eq:phi_sum}{eq:Apar_sum} reduce 
straightforwardly to the standard Reduced MHD equations.\cite{strauss:134} 
In the intermediate limit $\kperp\rho_i\gg1$, $\kperp d_e\ll1$, they 
reduce to the low-beta limit of the Electron Reduced MHD equations,\cite{alex} 
and so can support kinetic Alfv\'en waves. More generally, we 
show in \apref{app:lin} how the 
full collisionless gyrokinetic dispersion relation 
in the asymptotic limit of $\kperp\rho_e\ll1$ and low beta\cite{howes} 
is recovered from \eqsdash{eq:phi_sum}{eq:ge_sum}.
Finally, for the linear tearing mode, the paradigmatic linear problem of magnetic 
reconnection theory, 
\eqsdash{eq:phi_sum}{eq:ge_sum} recover the correct kinetic formulation 
both for the collisionless and the semicollisional regimes,\cite{cowley:3230} 
--- this topic is also treated in detail in \apref{app:lin}.

\section{Energetics\label{sec:energetics}}

\subsection{Free energy\label{sec:free}}

A broad class of $\delta f$ kinetic systems, including gyrokinetics, 
conserve (in the absence of collisions) 
a positive-definite quadratic quantity that has the physical meaning 
of the free energy of the combined system of particles and perturbed  
fields and plays the role of the generalized energy invariant
(see Refs.~\onlinecite{PPCFalex,alex} and references therein):
\bea
\nonumber
W &=& \sum_s\int\frac{d^3\vx}{V}\int d^3\vv\,\frac{T_{0s}\delta f_s^2}{2F_{0s}}
+ \int\frac{d^3\vx}{V}\frac{|\delta \mathbf{B}|^2}{8\pi}\\
&=& - \sum_s T_{0s}\delta S_s + \Um, 
\label{eq:Wgen}
\eea 
where $\delta S_s$ is the perturbed entropy of species $s$
and $\Um$ is the energy of the perturbed magnetic field.
Since energy flows play a fundamental role in all nonlinear 
phenomena, it will be instructive to understand the energetics 
of our equations. As KREHM is a partucular limit 
of gyrokinetics, this is done by specializing from the more general 
gyrokinetic case. 

We saw in \secref{sec:els} that  
\beq
\label{eq:dfe}
\delta f_e = \lt(\frac{\delta n_e}{n_{0e}} + \frac{2\vpar\upare}{\vthe^2}\rt)F_{0e} + g_e
\eeq
[see \eqref{eq:eansatz}]. 
Hence we find immediately that the electron perturbed entropy is
\bea
\nonumber
-T_{0e}\delta S_e &=& \int\frac{d^3\vx}{V}
\lt(\frac{n_{0e}T_{0e}}{2}\frac{\delta n_e^2}{n_{0e}^2}
+ \frac{m_e n_{0e}\upare^2}{2}\rt.\\
&& +\lt. \int d^3\vv\,\frac{T_{0e}g_e^2}{2F_{0e}}\rt).
\eea
The three terms here are the electron density variance 
(denoted $Y$ and, as we shall see, interpretable as 
generalized enstrophy), 
the kinetic energy of the parallel electron flow 
(denoted $\Ee$) and 
the free energy associated with the reduced electron distribution 
function $g_e$, which we will refer to as {\em electron free energy}
and denote $\We$. Using \eqsref{eq:Ampere}, \exref{eq:Poisson} 
and \exref{eq:upari}, we rewrite the above expression as follows
\begin{align}
\nonumber
-&T_{0e}\delta S_e = \Enst + \Ee + \We\\
\nonumber
&= \frac{n_{0e}T_{0e}}{2}\sum_{\vk}
\frac{Z^2}{\tau^2}\lt[1-\Gamma_0(\alpha_i)\rt]^2
\frac{e^2|\varphi_\vk|^2}{T_{0e}^2}\\
& + \int\frac{d^3\vx}{V} \frac{d_e^2|\nabla_\perp^2\Apar|^2}{8\pi}
+ \int\frac{d^3\vx}{V}\int d^3\vv\,\frac{T_{0e}g_e^2}{2F_{0e}},
\label{eq:Se}
\end{align}
where $\Gamma_0$ was defined in \eqref{eq:Gamma}. 

Using \eqref{eq:dfi} for $\delta f_i$, we find the ion perturbed entropy:
\beq
-T_{0i}\delta S_i = \frac{n_{0e}T_{0e}}{2}\sum_{\vk}
\frac{Z}{\tau}\lt[1-\Gamma_0(\alpha_i)\rt]
\frac{e^2|\varphi_\vk|^2}{T_{0e}^2}\equiv\Ei.
\label{eq:Si}
\eeq
Note that, were this quantity restricted to $\kperp\rho_i\ll1$, 
it would simply be the kinetic energy of the 
$\mathbf{E}\times\mathbf{B}$ flows: indeed, using 
\eqref{eq:Gamma_lw} and assuming an expansion in $\tau$ (as explained 
just after the latter formula), 
the quantity given by \eqref{eq:Si} becomes
\beq
\Ei=\int\frac{d^3\vx}{V}\frac{e^2 n_{0e}}{2T_{0e}}\,\rhos^2|\vdel_\perp\varphi|^2
= \int\frac{d^3\vx}{V}\frac{m_in_{0i}\uperp^2}{2},
\eeq
where $\uperp=c|\vdel_\perp\varphi|/B_0$ is the $\mathbf{E}\times\mathbf{B}$ 
flow velocity. Under the same approximation, the first term on the 
right-hand side of \eqref{eq:Se} (which arose from the electron 
density variance) is recognizable as the enstrophy of 
the $\mathbf{E}\times\mathbf{B}$ flow:
\beq
\Enst=\int\frac{d^3\vx}{V}\frac{e^2 n_{0e}}{2T_{0e}}\,\rhos^4|\nabla_\perp^2\varphi|^2.
\eeq

Finally, the magnetic energy is
\beq
\Um = \int\frac{d^3\vx}{V} \frac{\dBperp^2}{8\pi} 
= \int\frac{d^3\vx}{V} \frac{|\vdel_\perp\Apar|^2}{8\pi} 
\label{eq:U}
\eeq
because $\dBpar$ is subdominant under our ordering [\eqref{eq:dBpar}]. 
Combining \eqsref{eq:Se}, \exref{eq:Si} and \exref{eq:U} to reassemble 
the total free energy [\eqref{eq:Wgen}], we get
\bea
\nonumber
W &=& \Ei + \Enst + \Um + \Ee + \We \\
\nonumber
&=&\sum_\vk
\lt[1+\frac{Z}{\tau}\lt(1-\Gamma_0\rt)\rt]
\frac{Z}{\tau}\lt(1-\Gamma_0\rt)
\frac{e^2 n_{0e}|\varphi_\vk|^2}{2T_{0e}}\\
\nonumber
&&+\int\frac{d^3\vx}{V} \frac{|\vdel_\perp\Apar|^2 + d_e^2|\nabla_\perp^2\Apar|^2}{8\pi}\\
&&+\int\frac{d^3\vx}{V}\int d^3\vv\,\frac{T_{0e}g_e^2}{2F_{0e}}.
\eea

In three dimensions (3D), $W$ is the only quadratic invariant of our equations. 
In two dimensions (2D), there is an additional family of invariants (inherited from 
the more general 2D invariants of gyrokinetics\cite{alex}) --- they are worked out 
in \apref{app:2D}. Their existence, while opening interesting avenues of investigation
of an academic kind, suggests that one should be very cautious in 
generalizing 2D analytical and numerical results to 3D reality (similarly to 
the situation in fluid turbulence theory, where nonlinear interactions and energy 
flows are dramatically different in 2D and in 3D).  

\subsection{Energy exchange between fields\label{sec:enex}}

It is illuminating to consider separately the time evolution 
of the five constituent parts of $W$. Multiplying \eqref{eq:phi_sum}
by $e\varphi/T_{0e}$ or by $(Z/\tau)(1-\hat{\Gamma}_0)e\varphi/T_{0e}$ and integrating 
over space, we obtain the evolution of the ion kinetic energy and enstrophy, respectively:
\bea
\nonumber
\frac{d\Ei}{dt} &=& 
\frac{c}{4\pi}\int\frac{d^3\vx}{V}\,\varphi\,\vb\cdot\vdel\nabla_\perp^2\Apar\\
\label{eq:Ei}
&=& -\int\frac{d^3\vx}{V}\,\Epar\jpar,\\
\frac{d\Enst}{dt} &=& 
\int\frac{d^3\vx}{V}\,\jpar\vb\cdot\vdel\lt[\frac{Z}{\tau}\,(1-\hat{\Gamma}_0)\varphi\rt],
\eea
where $\Epar =-\vb\cdot\vdel\varphi$ is the electrostatic part of the parallel 
electric field and $\jpar=-(c/4\pi)\nabla_\perp^2\Apar$ is the parallel current. 
Multiplying \eqref{eq:Apar_sum} by $\nabla_\perp^2\Apar/4\pi$ and integrating over space, 
we obtain the evolution of the combined magnetic and electron kinetic energies: 
\begin{align}
\nonumber
&\frac{d}{dt}\lt(\Um + \Ee\rt) = \int\frac{d^3\vx}{V}\lt\{-\frac{4\pi}{c^2}\,\eta\jpar^2 
+ \Epar\jpar\rt.\\ 
&\lt.- \jpar\vb\cdot\vdel\lt[\frac{Z}{\tau}\,(1-\hat{\Gamma}_0)\varphi\rt]
+\frac{1}{e}\,\jpar\vb\cdot\vdel\dTe\rt\}.
\label{eq:mag}
\end{align}

With the help of \eqsdash{eq:Ei}{eq:mag}, we now examine the evolution of the combined 
``fluid'' (electromagnetic) part of the free energy, $\Wf = \Ei+\Enst+\Um+\Ee$, 
which is the quantity 
that is normally considered to be conserved in two-fluid models of Hamiltonian 
collisionless reconnection\citep{schep-phpl-94-twofluid}. 
The energy exchange terms containing $\varphi$ cancel and we find that $\Wf$ can only change 
due to two effects: the resistive Ohmic dissipation [the first term on the right-hand 
side of \eqref{eq:mag}],
and an exchange with the electron free energy $\We$ 
controlled by the last term on the right-hand side of \eqref{eq:mag}. 
Since $\jpar=-e n_{0e}\upare$ [see \eqsand{eq:Ampere}{eq:upari}], 
we can interpret this term as work done by 
the parallel electron pressure $\dpe=n_{0e}\dTe$: 
\beq
\int\frac{d^3\vx}{V}\frac{1}{e}\,\jpar\vb\cdot\vdel\dTe
= - \int\frac{d^3\vx}{V}\,\upare\vb\cdot\vdel\dpe \equiv - Q. 
\label{eq:Q_def}
\eeq
Thus, the fluid part of the free energy evolves according to
\beq
\frac{d}{dt}\,\Wf = -Q
-\frac{4\pi}{c^2}\,\eta\int\frac{d^3\vx}{V}\,\jpar^2.
\label{eq:Wf}
\eeq 
Note that $Q$ is not sign-definite and can correspond both 
to loss and gain of energy. In particular, it clearly represents 
a loss ($Q>0$) if the electron fluid is compressed. 

The fluid energy lost or gained this way is recovered in the evolution 
of the electron free energy, which we obtain by multiplying \eqref{eq:ge_sum} 
by $T_{0e}g_e/F_{0e}$ and integrating over the entire phase-space (positions and velocities):
\beq
\frac{d\We}{dt} = Q - \Dcoll,
\label{eq:We}
\eeq
where $\Dcoll$ is the dissipation term due to collisions, which must be positive 
definite for any collision operator that satisfies Boltzmann's $H$ theorem 
(see discussion in Ref.~\onlinecite{abel:122509} and references therein):
\beq 
\Dcoll = -\int\frac{d^3\vx}{V}\int d^3\vv\,\frac{T_{0e}g_eC[g_e]}{F_{0e}}\ge0.
\label{eq:Dcoll}
\eeq
Adding \eqsand{eq:Wf}{eq:We}, we find
that the total free energy $W$ is conserved in the absence of collisions,  
as stated at the beginning of this Section. 

\subsection{Dissipation, electron heating and irreversibility of 
collisionless reconnection\label{sec:heating}}

Consider some initial configuration prone to reconnection. 
Such a configuration will possess a certain amount of magnetic energy $\Um$, 
which in the process of reconnection will be converted into 
some other form. There can be two fundamentally different types of such 
energy conversion. 

First, since the conserved quantity is not $\Um$ but 
the total free energy $W=\Um + \Ee + \Ei + \Enst + \We$, 
the magnetic energy can be transferred dynamically into 
$\Ee$, $\Ei$, $\Enst$ or $\We$ without loss of $W$, 
or increase of entropy --- therefore, in principle, reversibly. 

Second, the magnetic energy can be 
dissipated via the resistive term in \eqref{eq:Wf} and/or 
via the collisional dissipation term 
$\Dcoll$ in \eqref{eq:We} --- in the latter case, it first 
has to be converted into the electron free energy via the energy 
exchange term $Q$ in \eqsand{eq:Wf}{eq:We}. These processes are 
irreversible because they involve conversion of the fluctuation energy 
into the thermal energy of the bulk electron distribution, or electron 
heating. Indeed, it is not hard to show\cite{howes,PPCFalex} 
that 
\bea
\nonumber
\frac{3}{2}\,n_{0e}\,\frac{dT_{0e}}{dt}
&=&-\overline{\int\frac{d^3\vx}{V}\int d^3\vv\frac{T_{0e}\delta f_e}{F_{0e}}
\lt(\frac{\dd\delta f_e}{\dd t}\rt)_c}\\ 
&=& \overline{\Dcoll} 
+ \frac{4\pi}{c^2}\,\eta\int\frac{d^3\vx}{V}\,\overline{\jpar^2},
\eea
where \eqref{eq:dfe} was used to express $\delta f_e$ 
and overbar means averaging over dynamical timescales 
(in gyrokinetics, the rate of change of the mean equilibrium 
quantities is $\sim\eps^2\omega$, so the transport equations are 
obtained via intermediate time averaging\cite{howes,abel-tome}). 

If we formally forbade any collisions 
at all and set $\nu_{ei}=0$ exactly, the exchanges between different 
constituent parts of $W$ would be the only possible 
energy conversion mechanism. Indeed, Hamiltonian theories of collisionless 
reconnection based on conservative fluid models ($g_e=0$, $\Wf=\const$) 
find that the nonlinear stage of a reconnecting mode corresponds to 
a transfer of magnetic energy $\Um$ into ordered parallel electron 
kinetic energy $\Ee$, perpendicular ion kinetic energy $\Ei$, and enstrophy of 
the $\mathbf E \times \mathbf B$ flow $\Enst$ . In this situation, the only way 
energy can be lost is via ejection of 
material --- although the reconnection process remains technically
reversible.

Is this, however, a good approximation of what will in fact occur? 
We believe that {\em a priori} it is not because, 
even if the collisionality of the plasma is small, 
the dissipation terms cannot be neglected as large spatial gradients will 
imply finite currents and large velocity-space gradients 
will imply finite values of $C[g_e]$. The latter effect is especially important. 
Indeed, if resistivity (Joule heating) is ignored, magnetic energy can still 
be converted into electron heat via transfer into $\We$, formation 
of small-scale structure in $\vpar$ by means of 
linear phase-mixing (the second term on the left-hand side in \eqref{eq:ge_sum}; 
see further discussion in \secref{sec:phmix}) 
and the consequent collisional dissipation 
of the resulting fine-scale electron distribution: since the collision 
operator is a diffusion operator in $\vpar$ [see \eqref{eq:C_LB}], 
$\Dcoll$ will have a finite value provided structure develops in 
the velocity-space that satisfies 
\beq
\nu_{ei}\frac{\dd^2}{\dd\hvpar^2}\gtrsim\omega
\quad\Rightarrow\quad
\frac{\delta\vpar}{\vthe}\lesssim \lt(\frac{\nu_{ei}}{\omega}\rt)^{1/2}.
\eeq
This phase-mixing channel of electron heating is open even in very 
weakly collisional plasmas. It seems {\em a priori} clear that 
its presence should be qualitatively important as it breaks the 
Hamiltonian nature of the problem and renders 
the process of ``collisionless'' reconnection irreversible.

The question of course remains whether the phase mixing will 
be important in any given physical situation of interest. 
Since most such situations are nonlinear, it is difficult 
to provide a universal answer to this question beyond the 
general argument that nature rarely ignores an energy dissipation 
channel if one is available. 
Here it is opportune to remind the reader 
that there have been several recent numerical studies of nonlinear 
kinetic reconnection that reported the non-negligibility 
of the parallel electron temperature gradient term in the 
generalized Ohm's law\cite{bowers:035002,daughton:072117,perona:042104,numata} 
[see \eqref{eq:Apar_sum}]
--- and, therefore, the non-negligibility of the energy exchange term $Q$ 
in \eqref{eq:Wf}, which ultimately transfers free energy into electron heat
[\eqref{eq:We}]. It is a testable prediction that in all such cases, small-scale 
structure should be generated in the velocity space (see \secref{sec:Hspectrum}), 
so as to enable the electron heating channel. 

In \secref{sec:phmix}, we discuss 
the electron kinetics a little further, in particular quantifying
the notion of the free-energy transfer to small parallel scales in 
velocity space. We will conclude (in \secref{sec:heatingrate}) that 
for configurations that evolve sufficiently slowly in time, 
the electron heating rate is finite and independent of the collision 
frequency as $\nu_{ei}\to+0$. 

\section{Velocity-space dynamics\label{sec:phmix}}

\subsection{Hermite expansion}

The velocity-space dynamics and the emergence of small-scale structure  
in $\vpar$ are best understood in terms of the expansion of 
the electron distribution function $g_e$ in Hermite 
polynomials.\cite{grant-feix1,armstrong,hammett-hermite,parker-carati,sugama:2617} 
Let 
\beq
g_e(\vpar) = \sum_{m=0}^\infty \frac{H_m(\hvpar)}{\sqrt{2^m m!}}\,\hg_m F_{0e}(\vpar),
\label{eq:ge_exp}
\eeq
where $\hvpar=\vpar/\vthe$, perpendicular velocity dependence is understood to 
have been integrated out, the Hermite polynomials are 
\beq
H_m(\hvpar) = (-1)^m e^{\hvpar^2}\frac{d^m}{d\hvpar^m}\,e^{-\hvpar^2}
\eeq 
(so $H_0=1$, $H_1=2\hvpar$, $H_2=4\hvpar^2-2$, etc.) and  
$\hg_m$ are the Hermite expansion coefficients (dimensionless), 
which can be calculated according to
\beq
\hg_m = \frac{1}{n_{0e}}\int_{-\infty}^{+\infty} d\vpar\,
\frac{H_m(\hvpar)}{\sqrt{2^m m!}}\,g_e(\vpar).
\label{eq:Htransform}
\eeq
In view of \eqref{eq:properties}, we must have $\hg_0=\hg_1=0$. 
\Eqref{eq:dTe} implies, by definition, that 
\beq
\hg_2 = \frac{1}{\sqrt{2}}\frac{\dTe}{T_{0e}}.
\label{eq:g2}
\eeq
Using the orthogonality of Hermite polynomials, 
\beq
\frac{1}{n_{0e}}\int_{-\infty}^{+\infty} d\vpar\,
\frac{H_m(\hvpar)H_n(\hvpar)}{2^mm!}\,F_{0e} = \delta_{mn},
\label{eq:orth}
\eeq
we note that the electron free energy in terms of the Hermite coefficients is 
\bea
\nonumber
\We &=& \int\frac{d^3\vx}{V}
\int_{-\infty}^{+\infty} d\vpar\,\frac{T_{0e} g_e^2}{2F_{0e}}\\
&=& \int\frac{d^3\vx}{V}\frac{n_{0e}T_{0e}}{2}\sum_{m=2}^{\infty}\hg_m^2. 
\label{eq:We_gm}
\eea

Taking the Hermite transform [\eqref{eq:Htransform}] of \eqref{eq:ge_sum}
and using the recursive property of the Hermite polynomials, 
\beq
\hvpar H_m(\hvpar) = \frac{1}{2}\,H_{m+1}(\hvpar) + m H_{m-1}(\hvpar),
\eeq
we arrive at
\begin{align}
\nonumber
&\frac{d\hg_m}{dt}\\ 
\nonumber
&+\vthe\vb\cdot\vdel\lt(\sqrt{\frac{m+1}{2}}\,\hg_{m+1} 
+ \sqrt{\frac{m}{2}}\,\hg_{m-1} - \delta_{m,1}\,\hg_2\rt)\\ 
\nonumber
&= - \sqrt{2}\,\delta_{m,2}\vb\cdot\vdel\frac{e}{cm_e}\,d_e^2\nabla_\perp^2\Apar\\ 
&\quad -\nu_{ei} \lt(m\hg_m - 2 \delta_{m,2}\,\hg_2\rt).
\label{eq:gm_gen}
\end{align}
Note that Hermite polynomials are eigenfunctions of the Lenard--Bernstein 
operator \exref{eq:C_LB}. 

It is not hard to see that \eqref{eq:gm_gen} is consistent with 
$\hg_0=\hg_1=0$. We would like to recast the equation for $\hg_2$ 
in terms of the electron temperature perturbation with the aid 
of \eqref{eq:g2}: 
\beq
\frac{d}{dt}\frac{\dTe}{T_{0e}} + \vthe\vb\cdot\vdel\sqrt{3}\,\hg_3
= - 2\vb\cdot\vdel \frac{e}{cm_e}\,d_e^2\nabla_\perp^2\Apar.
\label{eq:dTe_ev}
\eeq 
This equation is coupled to the rest of the kinetics via 
the heat flux proportional to $\hg_3$ (the second term on 
the left-hand side). For $m\ge3$, we have
\begin{align}
\nonumber
\frac{d\hg_m}{dt} &+ \vthe\vb\cdot\vdel\lt(\sqrt{\frac{m+1}{2}}\,\hg_{m+1} 
+ \sqrt{\frac{m}{2}}\,\hg_{m-1}\rt)\\
&= -\nu_{ei} m\hg_m.
\label{eq:gm}
\end{align}
\Eqref{eq:gm} shows that Hermite modes of order $m$ are coupled both 
to lower ($m-1$) and higher ($m+1$) modes. 
Collisions provide a cutoff at sufficiently large $m$ regardless of 
the magnitude of the collision frequency. 
We will discuss this further in \secref{sec:Hspectrum}. 

Together with \eqsand{eq:phi_sum}{eq:Apar_sum}, 
\eqsand{eq:dTe_ev}{eq:gm} can be solved as a system of 
partial differential equations in the three- or two-dimensional position 
space. This appears to be an attractive way to carry out numerical 
simulations of collisionless (in fact, weakly collisional) reconnection
or, indeed, of other kinetic phenomena in strongly magnetized plasmas. 
This system is substantially simpler than 
the full kinetic\cite{ricci:4102,daughton:072101,drake:042306,daughton:065004,daughton:072117}
or gyrokinetic\cite{wan:012311,rogers:092110,wang:072103,perona:042104,numata}
descriptions that have been used to study reconnection
or plasma turbulence\cite{bowers:035002,howes:065004} so far
(see \secref{sec:Hspectrum} for estimates of how many Hermite modes 
must be kept for any given collisionality). 
The first numerical study using our equations 
is reported in Ref.~\onlinecite{loureiro-zocco}.

The Hermite formalism allows us to provide 
very concise derivations of three important results: 
the $H$ theorem for our collision operator (\secref{sec:Htheorem}), 
the so-called semicollisional limit of our equations, 
in which collisionality dominates and the parallel electron temperature 
is determined by a fluid equation (\secref{sec:semicoll}), 
the Hermite ``spectrum,''   
which quantifies the fine structure in the velocity space 
caused by the parallel phase mixing
(\secref{sec:Hspectrum}), and finally the 
electron heating rate (\secref{sec:heatingrate}).

\subsection{$H$ theorem\label{sec:Htheorem}}

As we mentioned in \secref{sec:colls}, a key requirement for choosing 
a model collision operator is that it satisfies Boltzmann's $H$ theorem, 
i.e., that collisional dissipation leads to increase of entropy. 
This means that we must have $\Dcoll\ge0$ [see \eqref{eq:Dcoll}]. 
Substituting \eqref{eq:ge_exp}, using the collision operator given by 
\eqref{eq:C_LB} and the orthogonality of Hermite polynomials
[\eqref{eq:orth}], we get
\begin{align}
\nonumber
\frac{1}{n_{0e}}&\int_{-\infty}^{+\infty} d\vpar\,\frac{g_e C[g_e]}{F_{0e}} 
= -\nu_{ei}\sum_{m=0}^{\infty} m\hg_m^2 + 2\nu_{ei}\hg_2\\ 
&= -\nu_{ei}\sum_{m=3}^{\infty} m\hg_m^2 \le 0,
\label{eq:Dcoll_gm}
\end{align}
so $\Dcoll\ge0$, q.e.d.

\subsection{Semicollisional limit\label{sec:semicoll}}

Consider the semicollisional limit, $\nu_{ei}\gg\omega$ and $\kpar\mfpe\ll1$
(``semicollisional'' because the perpendicular microscale effects associated with 
$\rho_s$ and $\rho_i$ are retained, although the electron inertia term in \eqref{eq:Apar_sum} 
must be neglected compared with the resistive term). 
From \eqref{eq:gm}, it is clear that in this limit, the Hermite coefficients 
get smaller with~$m$:
\beq
\frac{\hg_m}{\hg_{m-1}}\sim \frac{\kpar\vthe}{\sqrt{m}\,\nu_{ei}}
=\frac{\kpar\mfpe}{\sqrt{m}}\ll1.
\eeq
Therefore, \eqref{eq:gm} allows us to express higher-order coefficients 
in terms of the gradients of the lower-order ones and, in particular, 
$\hg_3$ in terms of $\hg_2$, i.e., the heat flux in term the temperature 
gradient: 
\beq
\hg_3 \simeq - \frac{1}{2\sqrt{3}}\frac{\vthe}{\nu_{ei}}\,
\vb\cdot\vdel\frac{\dTe}{T_{0e}}. 
\label{eq:g3}
\eeq
Substituting this into \eqref{eq:dTe_ev}, we obtain the standard 
equation for the electron temperature:
\beq
\frac{d}{dt}\frac{\dTe}{T_{0e}} = \kape\vb\cdot\vdel\lt(\vb\cdot\vdel\frac{\dTe}{T_{0e}}\rt)
- 2\vb\cdot\vdel \frac{e}{cm_e}\,d_e^2\nabla_\perp^2\Apar.
\label{eq:dTe_Sp}
\eeq
where $\kape=\vthe^2/2\nu_{ei}$ is the parallel (Spitzer) thermal diffusivity. 
In the semicollisional limit, \eqref{eq:dTe_Sp} replaces \eqref{eq:ge_sum} 
and completes what has become a purely fluid system. 

The energetics of this fluid system are a simple particular case of the 
general situation discussed in \secref{sec:energetics}. The electron 
free energy is [see \eqref{eq:We_gm}]
\beq
\We = \int\frac{d^3\vx}{V}\frac{n_{0e}T_{0e}}{2}\,\hg_2^2 
= \int\frac{d^3\vx}{V}\frac{n_{0e}T_{0e}}{4}\frac{\dTe^2}{T_{0e}^2}. 
\eeq
From \eqref{eq:dTe_Sp}, we get an evolution equation for $\We$ 
equivalent to the general \eqref{eq:We}:
\beq
\frac{d\We}{dt} = Q 
- \kape\int\frac{d^3\vx}{V}\,\frac{n_{0e}T_{0e}}{2}
\lt(\vb\cdot\vdel\frac{\dTe}{T_{0e}}\rt)^2,
\eeq
where $Q$ is the energy exchange given by \eqref{eq:Q_def} 
and the dissipation term is the same as $\Dcoll$ defined by \eqref{eq:Dcoll}
--- this is checked by using \eqref{eq:Dcoll_gm} and restricting 
the sum just to $\hg_3$, as given by \eqref{eq:g3}.

\subsection{Hermite spectrum\label{sec:Hspectrum}}

We now return to \eqref{eq:gm}, linearize it and Fourier transform 
in the parallel direction. If we denote 
$\tg_m(\kpar) = (i\,\mathrm{sgn}\kpar)^m\hg_m(\kpar)$, we can write 
the resulting equation in the following form
\begin{align}
\nonumber
\frac{\dd\tg_m}{\dd t} &+ |\kpar|\vthe\lt(\sqrt{\frac{m+1}{2}}\,\tg_{m+1} 
- \sqrt{\frac{m}{2}}\,\tg_{m-1}\rt)\\
&= -\nu_{ei} m\tg_m,
\label{eq:tgm}
\end{align}
which has the convenient property of supporting real solutions.\footnote{It can 
also be thought to describe the nonlinear situation, rather than 
the linearized one, if \eqref{eq:gm} is first transformed into the Lagragian 
frame moving with the plasma and the parallel coordinate measured along the 
exact field line. This, however, is not an essential point as the phase mixing 
mechanism we are considering is fundamentally linear.} 
If we now define the Hermite spectrum as $E_m = |\hg_m|^2/2 = \tg_m^2/2$, 
we find that it evolves according to
\beq
\frac{\dd E_m}{\dd t} = - \lt(\flux_{m+1/2} - \flux_{m-1/2}\rt) 
- 2\nu_{ei} m E_m,
\eeq
where $\flux_{m-1/2} = |\kpar|\vthe\sqrt{m/2}\,\tg_m\tg_{m-1}$ can be 
thought of as the flux of the electron free energy through the Hermite 
space. When $m\gg1$, we may approximate 
$\flux_m\approx|\kpar|\vthe\sqrt{2m}\,E_m$ and
\beq
\frac{\dd E_m}{\dd t} = - |\kpar|\vthe\,\frac{\dd}{\dd m}\sqrt{2m}\,E_m 
- 2\nu_{ei} m E_m.
\label{eq:mflux}
\eeq

\subsubsection{Hermite spectrum in steady state}

\Eqref{eq:mflux} has a steady-state solution\footnote{The spectrum with the same $m$ 
dependence as \exref{eq:Em} was first derived in Ref.~\onlinecite{watanabe:1476} 
for electrostatic ITG turbulence in tokamaks, 
although, unlike those authors, we have not invoked any nonlinear 
dynamics in its derivation.} 
\beq
E_m = \frac{C(\kpar)}{\sqrt{m}}\,\exp\lt[-\lt(\frac{m}{\mc}\rt)^{3/2}\rt],
\label{eq:Em}
\eeq
where $C(\kpar)$ is some function of $\kpar$ (determined by the 
dynamics of $\varphi$, $\Apar$ and $\dTe$) 
and the collisional cutoff is
\beq
\mc = \lt(\frac{3}{2\sqrt{2}}\frac{|\kpar|\vthe}{\nu_{ei}}\rt)^{2/3}. 
\label{eq:mc}
\eeq
When $m\ll\mc$, we have the spectrum scaling as $E_m\sim m^{-1/2}$ 
in Hermite space. This scaling implies that the electron free energy 
$\We \propto \sum_{m=2}^\infty E_m$ is dominated by the Hermite modes
$m\sim\mc$, which are transferring their energy into electron heat via 
collisions (see \secref{sec:heatingrate}). 

\subsubsection{Hermite spectrum for growing modes}

Consider now a slightly more general situation, in which the entire 
Hermite spectrum is growing at some rate $2\gamma$ (this would be the case, 
for example, for the tearing mode; see \apref{app:tearing}). 
Then $\dd E_m/\dd t = 2\gamma E_m$ and the solution of \eqref{eq:mflux} is
\beq
E_m = \frac{C(\kpar)}{\sqrt{m}}\,\exp\lt[-\lt(\frac{m}{\mg}\rt)^{1/2}
-\lt(\frac{m}{\mc}\rt)^{3/2}\rt],
\label{eq:Em_growing}
\eeq
where another cutoff has appeared, associated with the growth rate $\gamma$: 
\beq
\mg = \lt(\frac{|\kpar|\vthe}{2\sqrt{2}\,\gamma}\rt)^2.
\eeq
This cutoff supercedes the collisional cutoff \exref{eq:mc} if 
$\mg\ll\mc$, i.e., if the mode is growing so fast that 
\beq
\gamma\gg (|\kpar|\vthe)^{2/3}\nu_{ei}^{1/3}. 
\label{eq:fastgrowth}
\eeq
Otherwise the steady-state spectrum \exref{eq:Em} is recovered. 

\subsection{Electron heating rate\label{sec:heatingrate}}

The results of \secref{sec:Hspectrum} allow us to make an estimate 
of the electron collisionless heating rate. Using \eqsand{eq:Dcoll}{eq:Dcoll_gm}, 
we may write 
\beq
\Dcoll = -n_{0e}T_{0e}\sum_{\kpar}2\nu_{ei}\sum_{m=3} mE_m.
\eeq
Approximating $\sum_{m=3} mE_m\approx \int_0^\infty dm\,mE_m$ 
and using \eqref{eq:Em}, we find 
for the steady-state or slowly evolving modes, 
\beq
\Dcoll \approx -n_{0e}T_{0e}\sum_{\kpar} \sqrt{2}\,|\kpar|\vthe C(\kpar) 
\eeq
{\em independently of the collision frequency} $\nu_{ei}$ as long as 
$\mc\gg1$, i.e., $\nu_{ei}\ll|\kpar|\vthe$. Thus, the electron heating rate 
is finite in the limit $\nu_{ei}\to+0$. This is analogous, e.g., to the 
situation in standard hydrodynamic turbulence where the rate 
of viscous dissipation has a finite limit at small viscosities 
because larger spatial gradients emerge\cite{K41} --- or in sub-Larmor gyrokinetic 
turbulence, where a similar process involves large gradients both 
in space and in $\vperp$.\cite{PPCFalex,alex,tatsuno:015003}  
The difference is that here the mechanism for the generation of small 
scales in $\vpar$ is linear (parallel particle streaming)  
rather than nonlinear (turbulent cascade). 

In contrast, for modes growing fast enough to satisfy \eqref{eq:fastgrowth}, 
a similar calculation using \eqref{eq:Em_growing} gives
\beq
\Dcoll \approx -n_{0e}T_{0e}\sum_{\kpar} \frac{\nu_{ei}}{2\sqrt{2}}
\lt(\frac{|\kpar|\vthe}{\gamma}\rt)^3C(\kpar). 
\eeq
In the limit $\nu_{ei}\to+0$, $\Dcoll\to0$. 
Thus, linearly growing modes do not produce any electron heating in the 
collisionless limit, provided the growth rate continues to satisfy \eqref{eq:fastgrowth} 
in this limit. 

\section{Discussion\label{sec:concs}}

In the above, we have already provided the motivation behind this work 
as well as a non-mathematical preview (\secref{sec:intro}), 
a summary of our equations (\secref{sec:summary}) 
and the general arguments leading to the conclusion that electron 
heating must be an important dissipation channel (\secref{sec:heating}). 
We developed this last topic somewhat further in \secref{sec:phmix} 
by setting forth what we believe to be a rather useful practical prescription 
for further numerical investigations, a conceptual view of the 
phase-space dynamics as a cascade in Hermite space, and a simple 
argument implying that the electron heating rate remains finite 
in the limit of vanishing collisionality (except for fast-growing 
linear modes). 

We do not of course claim that our equations constitute 
a general theoretical framework for all types of magnetic reconnection. 
The assumption of low beta restricts their applicability 
only to very strongly magnetized plasmas
(in \secref{sec:spscales}, we indicated that typical 
real-world plasmas where beta is suitably low 
include the solar corona,\cite{uzdensky:2139} 
the LArge Plasma Device at UCLA\cite{gekelman:2875} 
and edge regions in some tokamaks\cite{saibene:969}). 
We do, however, believe that at least qualitatively, 
our equations might go beyond their formal domain of validity 
and capture much of the essential physics of gyrokinetic 
reconnection. We do not know how to derive a similarly simple and yet physically 
realizable model for high-beta and low-guide-field situations, which occur 
frequently in astrophysical contexts. 

An unresolved limitation of our model concerns the problem of 
ultrasmall perpendicular scales. It has been found in several (2D)
theoretical and numerical studies of various two-fluid models of collisionless reconnection 
that sub-$d_e$ structures can form in the reconnection region, giving rise 
to unbounded gradients.\cite{drake-kleva,ottaviani-porcelli,valori:178,zocco:110703}  
The only known way of taming these singularities 
is via hyperdiffusive terms in the generalized Ohm's law, provided, e.g., 
by electron viscosity. Since we do not have these terms in our equations, 
it is not guaranteed that the singularities can be dissipated via the electron 
heating channel.\footnote{Note, however, the claim in Ref.~\onlinecite{blank:309} 
that the inclusion of electron kinetics alleviates somewhat the problem 
of formation of singularities in their 2D Hamiltonian fluid-kinetic model.} 
If they are present, any practical numerical study of our 
equations will require additional hyperdiffusive regularization --- 
a somewhat embarassing solution, since we would ideally like to be able to attribute 
all field-line breaking and electron heating to physical mechanisms that 
are legitimate under our ordering assumptions. We should like to note, however, 
that the evidence of singularities stems from 2D analyses of Hamiltonian 
reconnection models. This means that the phase space of the system is constrained 
both by the Hamiltonian structure and by the presence of multitudinous additional 2D invariants
(see \apref{app:2D}), so it is not, in fact, inevitable that these singularities 
will still present in a 3D kinetic situation. 

It seems clear that the next logical step is a numerical investigation of the 
solutions of our model. The key physical questions are whether reconnection is 
indeed made irreversible by the electron heating channel, how much energy 
is converted into heat, whether the reconnection is fast and what the structure 
of the reconnecting region is (X point? stable current sheet? unstable current 
sheet?). It would also be important to find out whether the equations are well 
posed without additional perpendicular regularization --- i.e., if singularities 
form. All of these issues are likely to be tied up with the question of whether 
the 3D case behaves differently from 2D and is, therefore, the right case to 
investigate. Finally, from a technical point of view, our model requires fairly 
modest computational resources, so it may prove to be a nimble tool for scouting out 
the parameter space in preparation for the more computationally demanding 
gyrokinetic reconnection studies.\cite{loureiro-zocco,numata} 

\acknowledgments

We are very grateful to N.~Loureiro, who 
read the paper in manuscript and offered numerous comments and suggestions. 
It is a pleasure to thank also
I.~Abel, M.~Barnes, T.~Carter, J.~Connor, S.~Cowley, P.~Dellar, W.~Dorland, 
R.~J.~Hastie, R.~Numata, J.~Parker, F.~Parra, B.~Rogers 
and D.~Uzdensky for useful discussions. 
Some of these discussions took place during the programme 
``Gyrokinetics in Laboratory and Astrophysical Plasmas'' at the 
Isaac Newton Institute, Cambridge, whose hospitality we gratefully acknowledge. 
This work started when AZ was a visitor at Oxford 
supported in part by the Leverhulme Trust Network for Magnetized Plasma Turbulence. 
AZ was subsequently supported by a Culham Fusion Research Fellowship and 
an EFDA Fellowship. 
The views and opinions expressed herein do not necessarily reflect 
those of the European Commissioners. 
AAS was supported in part by an STFC Advanced Fellowship
and the STFC Astronomy Grant ST/F002505/2.

\appendix

\section{Two-dimensional invariants\label{app:2D}}

\subsection{Lagrangian conservation properties in 2D}

Whereas the free energy [\eqref{eq:Wgen}] is the only known invariant of 
gyrokinetics in three dimensions (3D), formally restricting consideration
to two dimensions (2D) introduces a host of new conservation 
properties. Indeed, consider the gyrokinetic equation \eqref{eq:g1} 
and with $\dd/\dd z=0$ and no collisions. It can then be written 
as an advection equation of a single scalar quantity:
\begin{align}
\label{eq:gk_2D}
&\frac{\dd \hh_{s}}{\dd t} + \frac{c}{B_{0}}\{ \la\chi\ra _{\vRs},\hh_{s} \} = 0,\\
& \hh_s = h_s - \frac{q_s\la\chi\ra_{\vRs}}{T_{0s}}\,F_{0s}.
\end{align}
This means that the volume integral of any function of $\hh_s$ is 
conserved:
\beq
\frac{\dd}{\dd t}\int\frac{d^3\vRs}{V}\, f(\hh_s) = 0.
\label{eq:2D_gen}
\eeq
Note that $\hh_s$ is a function of $\vpar$ and $\vperp$, so the above 
equation defines an infinite set of invariants both in the sense of 
the arbitrariness of the function $f$ and in the sense that the invariants 
are parametrized by the velocity variables. 

Recalling our solution for the ion distribution function (\secref{sec:ions}), 
we find immediately that $\hh_i=g_i=0$, the conservation law \eqref{eq:2D_gen} 
is satisfied trivially. 
For the electrons, using \eqsref{eq:chie}, \exref{eq:eansatz}, 
\exref{eq:upare}, \exref{eq:Poisson}, and \exref{eq:upari}, we find
\begin{align}
\nonumber
\hh_e = 
-\frac{e}{cm_e}\lt(\Apar-d_e^2\nabla_\perp^2\Apar\rt)\frac{2\vpar}{\vthe^2}\,F_{0e}\\
-\frac{Z}{\tau}(1-\hat{\Gamma}_0)\frac{e\varphi}{T_{0e}}\,F_{0e} 
+ g_e.
\label{eq:hh}
\end{align}
This quantity obeys the advection equation \exref{eq:gk_2D}, which becomes 
\beq
\frac{\dd \hh_{e}}{\dd t} + \frac{c}{B_{0}}\{\chi,\hh_{e}\} = 0,
\label{eq:hh_adv}
\eeq
where $\chi=\varphi - (\vpar/c)\Apar$ [see \eqref{eq:chie}]. 
Thus, $\hh_e$, parametrized by $\vpar$, is an infinite family 
of ``Lagrangian invariants,'' each with its own stream function $(c/B_0)\chi(\vpar)$.
This is the ``foliation'' of the electron distribution function 
discussed in Refs.~\onlinecite{liseikina:3535,pegoraro:243}. 

While (as argued in Refs.~\onlinecite{liseikina:3535,pegoraro:243}) 
this may be reminiscent of the 
Lagrangian properties of the two-fluid model,\cite{schep-phpl-94-twofluid} 
setting $g_e=0$ in \eqref{eq:hh} does not return the Lagrangian 
invariants enjoyed by that model --- those have to be derived 
separately and are particular to the isothermal-electrons closure 
(which is not a surprise because $g_e=0$ is not a solution of our equations).
For the record, in our notation, they are\cite{schep-phpl-94-twofluid}
\beq
A_\pm = \Apar - d_e^2\nabla_\perp^2\Apar 
\pm \frac{c\sqrt{2}}{\vthe}\frac{Z}{\tau}(1-\hat{\Gamma}_0)\varphi,
\eeq
advected by the effective potentials 
\beq
\varphi_\mp = \varphi \mp \frac{\vthe}{c\sqrt{2}}\,\Apar
\eeq
(this is equivalent to picking $\vpar=\pm\vthe/\sqrt{2}$ and $g_e=0$ in 
\eqsand{eq:hh}{eq:hh_adv}). Note that $c\sqrt{2}/\vthe = cd_e/v_A\rho_s$. 
It is perhaps useful to give the evolution 
equations for $A_\pm$ with three-dimensionality, Ohmic resistivity 
and nonisothermal electrons retained, to show how the Lagrangian 
invariants are broken by all of these effects: 
\begin{align}
\nonumber
\lt(\frac{\dd}{\dd t}\pm\frac{\vthe}{\sqrt{2}}\frac{\dd}{\dd z}\rt)A_\pm 
+ \frac{c}{B_0}\lt\{\varphi_\mp,A_\pm\rt\} =\\ -\,\,c\,\frac{\dd\varphi_\mp}{\dd z}
+ \eta\nabla_\perp^2\Apar + \vb\cdot\vdel\frac{c}{e}\,\dTe.
\end{align}
For a model with no collisions, isothermal electrons and cold ions, 
these equations reduce to those of Ref.~\onlinecite{borgogno:032309}.
In the latter reference, its authors point out that, while the Lagrangian invariance 
of $A_\pm$ is lost in their 3D system, there is a cross-helicity 
invariant that continues to be conserved. We see, however, that this 
conservation law does not survive the inclusion 
of nonisothermal electrons:
\begin{align}
\nonumber
&\frac{d}{dt}\int\frac{d^3\vx}{V}\frac{A_+^2-A_-^2}{2} =\\ 
&\frac{2\sqrt{2}\,c}{\vthe}\int\frac{d^3\vx}{V}
\lt(\eta\nabla_\perp^2\Apar + \vb\cdot\vdel\frac{c}{e}\,\dTe\rt)
\frac{Z}{\tau}(1-\hat{\Gamma}_0)\varphi,
\end{align}
another constraint on the evolution of the system that 
disappears in our treatment. 

\subsection{Quadratic 2D invariants}

It is usually the quadratic invariants that are of most practical 
interest in nonlinear dynamics (or at least they are the ones that have 
the most physically transparent consequences, e.g., the inverse cascades 
in 2D gyrokinetic turbulence\cite{plunk-jfm}). 
In gyrokinetics, specializing to the quadratic 
function $f$ in \eqref{eq:2D_gen}, gives one the following family 
of 2D gyrokinetic invariants:\cite{alex}
\beq
I_s = \int d^3\vv\int\frac{d^3\vRs}{V}\,\frac{T_{0s}}{2F_{0s}}
\lt(h_s - \frac{q_s\la\chi\ra_{\vRs}}{T_{0s}}\,F_{0s}\rt)^2.
\eeq
We have also now integrated over the velocity space to produce 
a conservation law that globally constrains the system in the phase space. 
Using \eqref{eq:hh} to work out the electron invariant, we get 
\bea
\nonumber
I_e &=& \Enst + \frac{1}{d_e^2}\,\Astuff
+ 2\Um + \Ee + \We\\
&=& W + \frac{1}{d_e^2}\,\Astuff + \Um - \Ei,
\label{eq:intI}
\eea
where the definitions of, and evolution equations for 
the total free energy $W$, enstrophy $\Enst$, 
magnetic energy $\Um$, electron kinetic energy $\Ee$, 
electron free energy $\We$ and ion kinetic energy $\Ei$ 
can be found in \secref{sec:free} and the quantity 
\beq
\Astuff = \int\frac{d^3\vx}{V}\frac{\Apar^2}{8\pi}, 
\eeq
does not have a commonly agreed name, but is sometimes 
referred to as ``$\Apar^2$-stuff''. It is conserved in standard 
2D magnetohydrodynamics (MHD). The 2D conservation law we are about to 
derive is a generalization of this MHD result. 

\Eqref{eq:intI} tells us that the electron 2D invariant $I_e$
contains the total free energy $W$, which is itself conserved 
both in 2D and in 3D. 
The remainder of $I_e$ is, therefore, a 2D invariant 
on its own. After straightforward algebra analogous to the calculations
in \secref{sec:enex}, we find the evolution equation for this 
quantity:
\begin{align}
\nonumber
\frac{d}{dt}&\lt[\Astuff + d_e^2(\Um - \Ei)\rt] = - \eta\Um\\
\nonumber
&+ \frac{c}{4\pi}\int\frac{d^3\vx}{V}\frac{\dd\Apar}{\dd z}
\lt\{\lt[1+\frac{Z}{\tau}(1-\hat{\Gamma}_0)\rt]\varphi \rt.\\
&-\lt. d_e^2\nabla_\perp^2\varphi - \frac{1}{e}\,\dTe\rt\}.
\end{align}
This is a 2D collisionless invariant because the last term on the 
right-hand side vanishes if (and, generally speaking, only if) $\dd/\dd z=0$.\\

We will not delve any further into the mathematical consequences of the 
2D conservation laws and instead limit ourselves to remarking 
that the presence of so many constraints on the dynamics particular 
to the exactly 2D case ought to make one beware of too much optimism 
about the relevance of 2D results and intuitions to 3D dynamics.

\section{Linear theory\label{app:lin}}

\subsection{Linearized equations\label{app:lin_eqns}}

Since we will want to use the linearized equations that are about to be derived 
for treating the tearing mode problem, we would like to do the linearization 
around an equilibrium containing both the guide field $\vB_0=B_0\vz$ and 
some in-plane field --- gyrokinetically speaking, this in-plane field 
is part of the small perturbation of the guide field. Thus, we let 
\beq
\vB = B_0\vz + \dB_y^{(0)}(x)\vy + \dvB^{(1)}, 
\eeq
where $\dB_\perp^{(1)}\ll\dB_y^{(0)}\ll B_0$. 
Therefore, $\Apar = \Apar^{(0)}(x) + \Apar^{(1)}$, where  
$\dB_y^{(0)} = -d\Apar^{(0)}/dx \equiv B_0 f(x)$ and 
$\dvB^{(1)} = -\vz\times\vdel\Apar^{(1)}$. 
The function $f(x)$ will contain all the information about the 
in-plane equilibrium. 
Examining \eqsdash{eq:phi_sum}{eq:ge_sum}, it is not hard to see 
that an equilibrium of this form can be maintained by letting 
$\varphi^{(0)}=0$, $g_e^{(0)}=0$ and adding an equilibrium parallel electric 
field to \eqref{eq:Apar_sum} to maintain the equilibrium 
magnetic field against the Ohmic resistivity: $cE_\parallel^{(0)}=\eta B_0 f'(x)$. 

Inserting all this into \eqsdash{eq:phi_sum}{eq:ge_sum}, 
dropping the superscripts on the perturbed quantities and Fourier 
transforming with respect to $y$, $z$ and time, we get
\usewidetext{
\begin{widetext}
\begin{align}
\label{eq:phi_lin}
&-\omega\frac{Z}{\tau}(1-\hat{\Gamma}_0)\frac{e\varphi}{T_{0e}}=
\frac{e}{cm_e}\,d_e^2\lt[\kpar(x)\nabla_\perp^2\Apar - k_yf''(x)\Apar\rt],\\
\label{eq:Apar_lin}
&-\omega\lt[\Apar-d_{e}^{2}\nabla_\perp^2\Apar\rt]
= -i\eta\nabla_\perp^2\Apar + k_y c d_e^2 f''(x)\varphi
- \kpar(x)c\lt\{\lt[1+\frac{Z}{\tau}(1-\hat{\Gamma}_0)\rt]\varphi 
- \frac{1}{e}\,\dTe\rt\},\\ 
& -\omega g_{e} + \kpar(x)\vpar 
\lt(g_{e}- \frac{\dTe}{T_{0e}}\,F_{0e}\rt) = - iC[g_e] + 
\lt(1-\frac{2\vpar^{2}}{\vthe^{2}}\rt)F_{0e}
\frac{e}{cm_e}\,d_e^2\lt[\kpar(x)\nabla_{\perp}^{2}\Apar - k_y f''(x)\Apar\rt],
\label{eq:ge_lin}
\end{align}
\end{widetext}
}
\nowidetext{
\begin{align}
\nonumber
&-\omega\frac{Z}{\tau}(1-\hat{\Gamma}_0)\frac{e\varphi}{T_{0e}}=
\frac{e}{cm_e}\,d_e^2\lt[\kpar(x)\nabla_\perp^2\Apar\rt.\\ 
\label{eq:phi_lin}
&\qquad\qquad\qquad\qquad\qquad\qquad\lt.- k_yf''(x)\Apar\rt],\\
\nonumber
&-\omega\lt[\Apar-d_{e}^{2}\nabla_\perp^2\Apar\rt]
= -i\eta\nabla_\perp^2\Apar + k_y c d_e^2 f''(x)\varphi\\
\label{eq:Apar_lin}
&\qquad- \kpar(x)c\lt\{\lt[1+\frac{Z}{\tau}(1-\hat{\Gamma}_0)\rt]\varphi 
- \frac{1}{e}\,\dTe\rt\},\\ 
\nonumber
& -\omega g_{e} + \kpar(x)\vpar 
\lt(g_{e}- \frac{\dTe}{T_{0e}}\,F_{0e}\rt) = -iC[g_e]\\
\nonumber
&\qquad+\lt(1-\frac{2\vpar^{2}}{\vthe^{2}}\rt)F_{0e}
\frac{e}{cm_e}\,d_e^2\lt[\kpar(x)\nabla_{\perp}^{2}\Apar\rt.\\ 
&\qquad\qquad\qquad\qquad\qquad\qquad\lt.- k_y f''(x)\Apar\rt],
\label{eq:ge_lin}
\end{align}
}
where $\kpar(x) = k_z + f(x) k_y$ and $\nabla_\perp^2=\dd_x^2 - k_y^2$.  

\subsubsection{Collisionless limit}

The simplest approach to linear theory in a kinetic plasma 
is to consider the purely collisionless case. Although, as we explained above, 
this is not really a good limit, even in a weakly collisional plasma, 
the linear results obtained in this limit are useful in that they will allow 
us to make contact with the existing theories. The collisionless limit is also 
a useful route to certain linear results (like Landau damping) that do in fact 
depend on infinitesimal amount of velocity-space dissipation. 

Thus, let us set $\eta=0$ and $C[g_e]=0$ in \eqsand{eq:Apar_lin}{eq:ge_lin}, 
respectively. If we now solve \eqref{eq:ge_lin} for $g_e$ explicitly and then 
integrate over velocity space according to \eqref{eq:dTe_sum} to obtain 
the parallel electron temperature perturbation, we get, after standard alegebra,
\begin{align}
\nonumber
&\frac{\dTe}{T_{0e}} = \frac{2}{|\kpar(x)|\vthe}
\frac{Z(\zeta(x)) + \zeta(x) Z'(\zeta(x))}{Z'(\zeta(x))}\\
&\qquad\times\frac{e}{cm_e}\,d_e^2
\lt[\kpar(x)\nabla_\perp^2\Apar - k_y f''(x)\Apar\rt],
\end{align}
where $\zeta(x) = \omega/|\kpar(x)|\vthe$, $Z(\zeta)$ is the plasma 
dispersion function\cite{Fried} (not to be confused with $Z$ in 
the ion charge $Ze$) and $Z'(\zeta) = -2[1+\zeta Z(\zeta)]$. 
All kinetic effects are wrapped up in the above expression for 
the parallel elecron temperature perturbation. 
Using \eqref{eq:phi_lin}, we can simplify it somewhat:
\begin{align}
\nonumber
&\frac{1}{e}\,\dTe =
-2\zeta\,
\frac{Z(\zeta) + \zeta Z'(\zeta)}{Z'(\zeta)}
\frac{Z}{\tau}(1-\hat{\Gamma}_0)\varphi\\
&\qquad= \lt[1-2\zeta^2 + \frac{2}{Z'(\zeta)}\rt]
\frac{Z}{\tau}(1-\hat{\Gamma}_0)\varphi.
\label{eq:dTe_lin}
\end{align}
This can now be substituted into \eqref{eq:Apar_lin}, 
whereupon \eqsand{eq:phi_lin}{eq:Apar_lin} form a closed set. 

Recalling the gyrokinetic Poisson equation \exref{eq:Poisson}, 
\eqref{eq:dTe_lin} can be interpreted as an 
equation of state with $\zeta$-dependent effective adiabatic exponent: 
\bea
\label{eq:state}
\frac{\dTe}{T_{0e}} &=& (\adex - 1)\frac{\delta n_e}{n_{0e}},\\
\adex &=& 2\lt[\zeta^2 - \frac{1}{Z'(\zeta)}\rt].
\label{eq:adex}
\eea
This notation will be useful in our treatment of the tearing mode. 

\subsubsection{Semicollisional limit}

In the semicollisional limit, we abandon \eqref{eq:ge_lin} and instead 
use the linearized \eqref{eq:dTe_Sp} to determine the parallel electron 
temperature perturbation:
\begin{align}
\nonumber
&\lt[\omega+i\kape\kpar^2(x)\rt]\frac{\dTe}{T_{0e}} =\\ 
&\qquad\frac{2e}{cm_e}\,d_e^2\lt[\kpar(x)\nabla_\perp^2\Apar - k_y f''(x)\Apar\rt].
\end{align}
Using \eqref{eq:phi_lin}, we rewrite this as 
\beq
\frac{1}{e}\,\dTe = -\frac{2\omega}{\omega + i\kape\kpar^2(x)}
\frac{Z}{\tau}(1-\hat{\Gamma}_0)\varphi.
\label{eq:dTe_Sp_lin}
\eeq
This can be substituted into \eqref{eq:Apar_lin}, where also the resistive 
term is now retained, while the electron inertia must be neglected in comparison 
(because $\nu_{ei}\gg\omega$). 

Recasting \eqref{eq:dTe_Sp_lin} in the form \exref{eq:state}, 
we find the effective adiabatic exponent in the semicollisional case:
\beq
\adex = 1 + \frac{2\omega}{\omega + i\kape\kpar^2(x)}.
\eeq

\subsection{Kinetic Alfv\'en waves\label{app:waves}}

\subsubsection{Collisionless limit}

Let us make a digression and ascertain that the equations we have derived 
contain the wave dynamics that they are expected to contain. 
Formally, if we assume $k_z\gg k_yf(x)$, $k_z\gg k_y d_e^2 f''(x)$ and 
$k_z\dd^2/\dd x^2 \gg k_yf''(x)$, we find ourselves in a homogeneous 
plasma (all terms containing $f(x)$ can be neglected and $\kpar=k_z$). 
This means that we can now also Fourier transform in $x$ and, after 
a few lines of standard algebra, we obtain from \eqsref{eq:phi_lin}, 
\exref{eq:Apar_lin} and \exref{eq:dTe_lin} the following dispersion 
relation: 
\beq
\lt[\zeta^2 - \frac{\tau}{Z}\frac{\kperp^2 d_e^2/2}{1-\Gamma_0(\kperp^2\rho_i^2/2)}\rt]
\lt[1+\zeta Z(\zeta)\rt] = \frac{1}{2}\,\kperp^2 d_e^2, 
\label{eq:waves}
\eeq
where $\kperp^2=k_x^2+k_y^2$. 
It is not hard to check that this agrees with the gyrokinetic 
dispersion relation at low beta derived in Ref.~\onlinecite{howes} 
[their Eq.~(D17)]. 

Looking for solutions with $\zeta = \omega/|\kpar|\vthe\ll1$ and using 
$Z(\zeta)\approx i\sqrt{\pi} - 2\zeta$, we find the well known dispersion relation 
for Alfv\'en waves (both MHD and kinetic): 
\beq
\omega = 
\pm \kpar v_A \kperp\rho_i
\sqrt{\frac{1}{2}\lt[\frac{Z}{\tau} + \frac{1}{1-\Gamma_0(\kperp^2\rho_i^2/2)}\rt]}. 
\label{eq:KAW}
\eeq
The MHD Alfv\'en waves, 
\beq
\omega = \pm\kpar v_A,
\eeq
are recovered for $\kperp\rho_i\ll1$ (in which case 
$\Gamma_0\approx 1 - \kperp^2\rho_i^2/2$), the kinetic Alfv\'en waves,
\beq
\omega = \pm \sqrt{\frac{1}{2}\lt(1+\frac{Z}{\tau}\rt)}\,\kpar v_A \kperp\rho_i,
\eeq 
for $\kperp\rho_i\gg1$ ($\Gamma_0\approx 0$). Finally, the 
damping rate is found from \eqref{eq:waves} as a small perturbation 
of $\omega$: 
\beq
\gamma = - |\kpar|v_A\,\frac{\kperp^2\rho_i^2}{4}
\sqrt{\pi\,\frac{m_e}{m_i}\frac{Z^3}{\tau^2\beta_e}}.
\eeq
(note that this is valid for any $\kperp\rho_i$). 
This again is a well known formula 
for the electron Landau damping of Alfv\'en waves at low beta
[cf.\ Eq.~(63) of Ref.~\onlinecite{howes} noting that 
$\beta_e=(Z/\tau)\beta_i$]. 

Another familiar and useful limit is the long-wavelength, cold-ion approximation, 
$\kperp\rho_i\sim\sqrt{\tau}\ll1$ (already discussed in \secref{sec:ions}). 
Using \eqref{eq:Gamma_lw} in \eqref{eq:waves}, we get
\beq
\lt(\zeta^2 - \frac{d_e^2}{2\rho_s^2}\rt)\lt[1+\zeta Z(\zeta)\rt] 
= \frac{1}{2}\,\kperp^2 d_e^2. 
\eeq
Assuming again $\zeta\ll1$, we get for the Alfv\'en waves in this limit:
\bea
\omega &=& \pm \kpar v_A \sqrt{1+\kperp^2\rho_s^2},\\
\gamma &=& - |\kpar|v_A\,\frac{\kperp^2\rho_s^2}{2}
\sqrt{\pi\,\frac{m_e}{m_i}\frac{Z}{\beta_e}}.
\eea

The dispersive kinetic Alfv\'en waves 
are believed to play a key role in enabling fast kinetic 
reconnection,\cite{rogers-prl-01-disp-waves,chacon:ep-plasma}
so it is important that they are fully retained in our equations. 

Note that assuming $\zeta\ll1$ in the above calculation meant that we 
effectively adopted an isothermal limit [see \eqref{eq:dTe_lin}]. 
Enforcing this requirement on the frequency [\eqref{eq:KAW}] also 
eliminates the electron scales: $\kperp d_e\ll1$ (in the cold-ion 
limit, this is modified to $d_e\ll\rho_s$). The non-isothermal 
effects will come in at the electron inertial scale 
together with the breaking of the flux conservation.   

\subsubsection{Semicollisional limit}

Finally, in the semicollisional limit, using \eqref{eq:dTe_Sp_lin} 
instead of \eqref{eq:dTe_lin}, we get the following dispersion relation:
\begin{align}
\nonumber
&\omega(\omega + i\eta\kperp^2) = \kpar^2 v_A^2\kperp^2\rho_i^2\\
&\quad\times\frac{1}{2}\lt[\frac{Z}{\tau}\frac{3\omega + i\kape\kpar^2}{\omega + i\kape\kpar^2} 
+ \frac{1}{1-\Gamma_0(\kperp^2\rho_i^2/2)}\rt].
\end{align}
The isothermal limit, \eqref{eq:KAW}, this time with resistive damping, 
is recovered for $\kape\kpar^2\gg\omega$, or, equivalently, 
$\kperp d_e \ll \kpar\mfpe$. 

\subsection{Collisionless tearing mode\label{app:tearing}}

Let us now consider the limit opposite to that which produced 
a homogeneous situation, namely, $k_z\ll k_y f(x)$, so 
the problem becomes effectively 2D and $\kpar(x)\approx k_y f(x)$. 
This is the tearing mode problem. Since we are expecting a 
(purely) growing mode, let $\omega=i\gamma$. 
In the collisionless limit, \eqsref{eq:phi_lin}, \exref{eq:Apar_lin} 
and \exref{eq:dTe_lin} can now be rearranged to read
\begin{align} 
\nonumber
&\frac{i\gamma}{k_yf(x)\vthe}\frac{Z}{\tau}(1-\hat{\Gamma}_0)\varphi =\\ 
\label{eq:phi_tm}
&\quad-\frac{1}{2}\,d_e^2\lt[\dd_x^2 - k_y^2 - \frac{f''(x)}{f(x)}\rt]
\frac{\vthe}{c}\,\Apar,\\
\nonumber
&\frac{i\gamma}{k_yf(x)\vthe}\lt(1-d_e^2\dd_x^2\rt)\frac{\vthe}{c}\,\Apar =\\ 
&\qquad\qquad\qquad\ \lt[1 + \adex\,\frac{Z}{\tau}(1-\hat{\Gamma}_0)\rt]\varphi,
\label{eq:Apar_tm}
\end{align}
where $G(\zeta)$ is the effective adiabatic exponent given by \eqref{eq:adex}, 
$\zeta(x) = i\gamma/|k_yf(x)|\vthe$, and  
we have neglected $d_e^2 k_y^2$ and $d_e^2f''(x)/f(x)$ compared to unity 
in \eqref{eq:Apar_tm}.

As in all tearing-mode calculations, these equations will 
turn out to have a boundary layer around $x=0$. The reconnection 
of the magnetic flux will happen inside this layer, while outside 
the dynamics will be essentially MHD. The difference between 
the standard MHD situation\cite{FKR} and what we have here 
is that there are two microscales in the problem: 
$\rho_s$ (or $\rho_i$) and $d_e$, so the boundary layer 
will in fact be made of two nested boundary layers. 
We will first deal with the outer solution, 
then give a set of {\it a priori} estimates of the relevant scales 
and of the growth rate of the mode that follow from the structure 
of the problem in the inner region. 

\subsubsection{Outer (MHD) region\label{app:outer}}

Here we seek the solution at the equilibrium scales $l$, assuming 
$l\sim [f'(x)/f(x)]^{-1} \sim k_y^{-1} \gg\rho_s\gg d_e$. 
Accordingly, we can neglect $d_e^2\dd_x^2\ll1$ and 
$(Z/\tau)(1-\hat{\Gamma}_0)\approx -\rho_s^2(\dd_x^2-k_y^2)\ll 1$. 
We are not assuming that $\zeta$ is either small or large. 
Under these approximations, \eqref{eq:Apar_tm} becomes simply
\beq
\frac{i\gamma}{k_yf(x)\vthe}\frac{\vthe}{c}\,\Apar = \varphi.
\eeq
Substituting this into \eqref{eq:phi_tm}, we get
\beq
\lt[\dd_x^2 - k_y^2 - \frac{f''(x)}{f(x)}\rt]\Apar
= \frac{2\rho_s^2}{d_e^2}\,\zeta(x)(\dd_x^2 - k_y^2)\zeta(x)\Apar.
\label{eq:Apar_outer}
\eeq
Now note that, since $\sqrt{2}\rho_s/d_e=\vthe/v_A$, we have 
\beq
\frac{\sqrt{2}\rho_s}{d_e}\,\zeta(x) = \frac{i\gamma}{|k_yf(x)|v_A} 
= \frac{i\gamma}{\omega_{Ay}},
\eeq
where $\omega_{Ay}=|k_yf(x)v_A| = |k_yv_{Ay}|$ is the in-plane Alfv\'en frequency. 
Assuming $\gamma\ll\omega_{Ay}$ (tearing mode is sub-Alfv\'enically slow), 
we may then neglect the right-hand side of \eqref{eq:Apar_outer} and so 
end up with
\beq
\dd_x^2\Apar = \lt[k_y^2 + \frac{f''(x)}{f(x)}\rt]\Apar. 
\eeq 
This is the standard MHD tearing mode outer-region equation. 
Its solution depends on the particular choice of $f(x)$, but we will 
not need its detailed form. Since $\dB_y = -\dd_x\Apar$ 
must reverse direction at $x=0$, the sign of $\dd_x\Apar$ will 
be different on either side of $x=0$, so the derivative will have 
a discontinuity. This is quantified by 
\beq
\Delta' = \frac{1}{\Apar(0)}\lt[\dd_x\Apar\rt]^{+0}_{-0}, 
\label{eq:Dprime}
\eeq
the tearing-mode instability parameter. 
It is a measure of the singularity developed in the current 
by the ideal-MHD outer solution as it approaches the reconnection 
layer. The singularity is resolved by corrections to the ideal 
Ohm's law (electron inertia in the case we are considering). 
The solution in the boundary layer around $x=0$ will 
have to be matched to the value of $\Delta'$. 

\subsubsection{Inner region: equations\label{app:inner}}

At scales of order $\rho_s$ or smaller, we may expand the in-plane equilibrium
profile: $f(x)\approx xf'(0) \equiv x/L_s$. Then $\zeta(x)\approx i\delta/|x|$, where 
\beq
\delta=\frac{\gamma}{k_y\vthe}\,L_s 
= \frac{\gamma}{k_y v_A}\frac{d_e}{\sqrt{2}\,\rho_s}\,L_s.
\eeq
Since $\dd_x^2\gg k_y^2, f''(x)/f(x)$ at these scales, 
\eqsand{eq:phi_tm}{eq:Apar_tm} become
\begin{align} 
\label{eq:phi_inner}
&\frac{\delta}{x}\frac{Z}{\tau}(1-\hat{\Gamma}_0)\tphi =
\frac{1}{2}\,d_e^2\dd_x^2\Apar,\\
&\frac{\delta}{x}\lt(1-d_e^2\dd_x^2\rt)\Apar =
\lt[1 + \adex\,\frac{Z}{\tau}(1-\hat{\Gamma}_0)\rt]\tphi,
\label{eq:Apar_inner}
\end{align}
where $\hat{\Gamma}_0 = \hat{\Gamma}_0(\rho_i^2\dd_x^2/2)$, 
$\tphi = -(c/\vthe)i\varphi$, 
and the effective adiabatic exponent $\adex$ is given
by \eqref{eq:adex}, which we now rewrite as 
\beq
\adex\lt(\frac{x}{\delta}\rt) = 
-2\lt[\frac{\delta^2}{x^2} + \frac{1}{Z'(i\delta/|x|)}\rt].
\label{eq:Gvsx}
\eeq
This function has limiting values $\adex(0)=3$, $\adex(\infty)=1$
(electrons are isothermal for $x\gg\delta$) and 
a smooth monotonic transition in between. 

In order to benchmark our theory against existing treatments, let us remark 
that \eqsand{eq:phi_inner}{eq:Apar_inner} can easily be manipulated into 
a form that coincides with the collisionless 
tearing mode equations derived by Cowley, Kulsrud and Hahm\cite{cowley:3230} 
directly from the Vlasov-Maxwell kinetics: 
\bea
\label{eq:Cowley_Epar}
\frac{x}{\delta}\lt(\Apar - \frac{x}{\delta}\,\tphi\rt)
\sigma\lt(\frac{x}{\delta}\rt) &=& \frac{Z}{\tau}(1-\hat{\Gamma}_0)\tphi,\\
\frac{1}{2}\frac{x}{\delta}\,d_e^2\dd_x^2\Apar 
&=& \frac{Z}{\tau}(1-\hat{\Gamma}_0)\tphi,\qquad\quad
\label{eq:Cowley_jpar}
\eea
where the ``scaled effective conductivity'' is 
\beq
\sigma\lt(\frac{x}{\delta}\rt) 
= \lt[2+\frac{x^2}{\delta^2}\,\adex\lt(\frac{x}{\delta}\rt)\rt]^{-1}\!\!\!
=-\frac{1}{2}\frac{\delta^2}{x^2}\,Z'\!\lt(\frac{i\delta}{|x|}\rt).
\eeq
\Eqsand{eq:Cowley_Epar}{eq:Cowley_jpar} are Eqs.~(26) and (25), respectively, 
of Ref.~\onlinecite{cowley:3230}, specialized to the case of no equilibrium 
density gradient\footnote{This reduction is accomplished by setting 
the electron density gradient ($1/a$) and, therefore, 
$\omega_*= k_y cT_{0e}/eB_0 a$ to zero in all of their equations except Eq.~(25), 
where $(\omega_*/\omega)^2/\beta_p= d_e^2/2\delta^2$, because the density gradient 
in $\omega_*$ cancels with the one in $\beta_p = (4\pi n_{0e}T_{0e}/B_0^2)L_s^2/a^2$.}
[note that in Ref.~\onlinecite{cowley:3230}, $\delta$ is defined in terms
of $\omega$, not $\gamma$, so to get exactly their equations, one must replace 
$\delta\to -i\delta$, $\sigma\to-\sigma$ and restore $\tphi = -(c/\vthe)i\varphi$]. 
These equations determine the parallel electric field [\eqref{eq:Cowley_Epar}]
and the parallel current [\eqref{eq:Cowley_jpar}] in terms of the ion response. 
This response is, in general, nonlocal, which leads to considerable technical 
difficulties.\cite{drake:1341,antonsen-coppi,cowley:3230,porcelli:425} 

\subsubsection{Mathematical structure of the problem\label{app:math}}

Let us introduce a rescaling of the spatial variable $\xi=x/\din$, 
where $\din$ reflects the width of the inner solution and will be 
determined later. \Eqsand{eq:Apar_inner}{eq:phi_inner} can be rewritten 
as follows
\begin{align}
\label{eq:Apar_xi}
&\frac{\Apar}{\xi} - \frac{\din}{\delta}\,\tphi = 
\frac{d_e^2}{\din^2}\lt(1+\frac{\din^2}{2\delta^2}\,\adex\xi^2\rt)\frac{\Apar''}{\xi},\\
&\frac{2\delta\din}{d_e^2}\frac{Z}{\tau}(1-\hat{\Gamma}_0)\tphi = \xi\Apar'',
\label{eq:phi_xi}
\end{align}
where primes denote derivatives with respect to $\xi$ and we have used 
\eqref{eq:phi_xi} to express the $(Z/\tau)(1-\hat{\Gamma}_0)\tphi$ term in 
\eqref{eq:Apar_inner} in terms of $\Apar''$. 

Let $\chi(\xi)=\xi\Apar' - \Apar = \xi^2(\Apar/\xi)'$. 
Then $\Apar''=\chi'/\xi$. 
Differentiating once \eqref{eq:Apar_xi} and expressing 
$\Apar$ in terms of $\chi$, we get 
\begin{align}
\frac{d}{d\xi}\lt(\frac{1}{\xi^2}+\frac{\din^2}{2\delta^2}\,\adex\rt)
\chi' = \frac{\din^2}{d_e^2}\frac{\chi}{\xi^2}  
- \frac{\din^3}{d_e^2\delta}\,\tphi'. 
\label{eq:chi_int}
\end{align}
We now need a closed expression for $\tphi$ in terms of $\Apar$, which 
is where the nonlocality of the ion response --- the operator 
on the left-hand side of \eqref{eq:phi_xi} --- presents a technical 
challenge. The simplest choice is to avoid it by considering the 
limit of cold ions. It turns out, however, that with no additional 
trouble, it is possible to accommodate (non-rigorously) the hot-ion 
limit by using the Pad\'e approximant for the ion response:\cite{pegoraro:647} 
\beq
\frac{Z}{\tau}(1-\hat{\Gamma}_0) \approx -\frac{\rho_s^2\dd_x^2}{1-(1/2)\rho_i^2\dd_x^2}.
\eeq
Then \eqref{eq:phi_xi} becomes
\beq
-\frac{2\delta\rho_s^2}{d_e^2\din}\,\tphi'' = \chi' - \frac{\rho_i^2}{2\din^2}\,\chi'''.
\eeq
By integrating this equation once, we work out the $\tphi'$ term in \eqref{eq:chi_int}:
\beq
-\frac{\din^3}{d_e^2\delta}\,\tphi' = \chi - \chi_0 
- \frac{\tau}{Z}\frac{\rho_s^2}{\din^2}\,\chi'',
\label{eq:phiprime}
\eeq
where $\chi_0$ is a constant of integration, we have used $\rho_i^2/2\rho_s^2 = \tau/Z$, 
and arranged for the overall coefficient in the left-hand side to be unity by choosing 
the inner rescaling width
\beq
\din = \lt(\sqrt{2}\,\rho_s\delta\rt)^{1/2}.
\label{eq:dindef}
\eeq
It remains to substitute \eqref{eq:phiprime} into \eqref{eq:chi_int} 
and arrange terms neatly. The result is
\begin{align}
\xi^2\frac{d}{d\xi}\lt[\frac{1}{\xi^2}+\alpha^2\lt(G + \frac{\tau}{Z}\rt)\rt]\tchi' 
- \lt(\xi^2 + \lambda^2\rt)\tchi = \lambda^2,
\label{eq:chi_eq}
\end{align}
where $\tchi=-1 + \chi/\chi_0$, 
\bea
\label{eq:ldef}
\lambda^2 &=& \frac{\din^2}{d_e^2} 
= \frac{\sqrt{2}\,\rho_s\delta}{d_e^2} = \frac{\gamma}{k_y v_A}\frac{L_s}{d_e},\\
\alpha &=& \frac{\din}{\sqrt{2}\,\delta} = 
\lt(\frac{\rho_s}{\sqrt{2}\,\delta}\rt)^{1/2}\!\!\! = \frac{\rho_s}{d_e\lambda},
\label{eq:adef} 
\eea
and $\adex(x/\delta)$ is given by \eqref{eq:Gvsx}, with $x/\delta=\sqrt{2}\,\alpha\xi$. 

The above equation is subject to matching with the outer solution. 
First, using the definition of $\chi$ and the property $\Apar''=\chi'/\xi$, 
we can write 
\beq
\Apar = -\chi_0 \lt[1 + \tchi - \xi\int_0^\xi dz\,\frac{\tchi'(z)}{z}\rt].
\label{eq:Apar_chi}
\eeq
This is matched to the outer (MHD) solution (see \apref{app:outer}) by 
identifying the limit $\xi\to\infty$ for the inner solution with 
$x\to0$ for the outer solution: this gives 
$\tchi(\xi\to\infty)=0$, $\chi_0=-\Apar(\xi\to\infty)=-\Apar(x\to0)$ 
and, from the condition \exref{eq:Dprime}, 
\beq
\int_0^\infty d\xi\,\frac{\tchi'(\xi)}{\xi} = -\frac{1}{2}\,\Delta'\din.
\label{eq:match}
\eeq 

\Eqsand{eq:chi_eq}{eq:match} constitute an eigenvalue problem, where 
$\lambda^2$ is the rescaled eigenvalue and $\alpha$ is the parameter 
that measures the width of the ion layer. 
Due to the double-layered structure of the inner region, 
the exact solution requires some 
work\cite{coppi-res,drake:2509,pegoraro:364,porcelli:425} 
(see \apref{app:twofluid_sln}), but all the relevant scalings 
can be derived in a simple heuristic way. While these 
scalings are known, it is useful to summarize them in a unified 
exposition. 

\subsubsection{Ultralow-beta (one-fluid) limit\label{app:ultralowbeta}}

If $\alpha\ll1$ ($\rho_s\ll\delta$), 
the ion effects are negligible and our equations turn into the standard 
tearing mode equations\cite{FKR} extended to arbitrary $\Delta'$\cite{coppi-res,ABC} 
and with resistivity replaced by electron inertia.\cite{laval-pellat-vuillemin,basu:465}

It is easy to derive the scaling of the growth rate without an exact 
solution of the equations. From \eqref{eq:chi_eq}, $\lambda^2\sim\tchi'/\xi$. 
The width of the $\tchi(\xi)$ function 
is order unity (i.e., $x\sim\din$ in dimensional terms), so 
the integral in \eqref{eq:match} is over an interval of order unity.
Therefore 
\beq
\lambda^2\sim \Delta'\din.
\label{eq:lambda_gen}
\eeq
Using the definitions \exref{eq:dindef} and \exref{eq:ldef}, 
one gets the scalings\cite{laval-pellat-vuillemin,drake:1341} 
\beq
\delta\sim\frac{d_e^4\Delta'^2}{\rho_s},\quad
\din\sim d_e^2\Delta',\quad
\gamma\sim k_y v_A\,\frac{d_e^3\Delta'^2}{L_s}.
\label{eq:lowb_fin}
\eeq
This result breaks down when $\Delta'$ is so large that 
$\Delta'\din\sim1$ or larger, in which case 
the magnitude of the current is limited by the 
width of the reconnecting region, $\din$. 
The correct scalings are obtained by replacing 
$\Delta'$ by $1/\din$ in \eqref{eq:lambda_gen}:
\beq
\lambda^2\sim1
\label{eq:lambda1}
\eeq 
(in other words, instead of estimating the current as 
$\dd_x^2\Apar\sim(\Delta'/\din)\Apar$, one takes 
$\dd_x^2\Apar\sim\Apar/\din^2$).
Hence one gets the scalings\cite{drake:1777,basu:465} 
\beq
\delta\sim\frac{d_e^2}{\rho_s},\quad
\din\sim d_e,\quad
\gamma\sim k_y v_A\,\frac{d_e}{L_s}.
\label{eq:lowb_inf}
\eeq

As we explained above, all of this is valid provided $\rho_s\ll\delta$, 
which amounts to 
\beq
\frac{\rho_s^2}{d_e^2} \sim \beta_e\,\frac{m_i}{m_e} \ll (d_e\Delta')^2 
\eeq
for the finite-$\Delta'$ scalings \exref{eq:lowb_fin} and
\beq
\frac{\rho_s^2}{d_e^2} \sim \beta_e\,\frac{m_i}{m_e} \ll 1
\label{eq:onefluid}
\eeq
for the infinite-$\Delta'$ limit \exref{eq:lowb_inf}. 
Either condition can only be satisfied if $\beta_e$ is very low indeed
--- not just of order, but much smaller than $m_e/m_i$
[obviously, this is treated as a subsidiary limit 
within our formal low-beta ordering \exref{eq:beta}].  

\subsubsection{Two-fluid case\label{app:twofluid}}

If $\alpha\gg1$, or $\rho_s\gg\delta$, 
the problem contains both ion and electron scales and is no 
longer mathematically equivalent to the resistive tearing mode
[the terms involving $\alpha$ in \eqref{eq:chi_eq} cannot be dropped]. 
The width of the integration region in \eqref{eq:match} is now 
$\xi\sim1/\alpha$ (or, dimensionally, $x\sim\delta$). 
Therefore,
\beq
\lambda^2\sim \Delta'\din\alpha.
\label{eq:lambdaalpha_gen}
\eeq
This gives immediately the following 
scalings\cite{porcelli:425,zakharov:3285,kuvshinov:867,mirnov:4468} 
\beq
\delta\sim d_e^2\Delta',\quad
\din\sim d_e \rho_s^{1/2}\Delta'^{1/2},\quad
\gamma\sim k_y v_A\,\frac{d_e\rho_s\Delta'}{L_s}.
\label{eq:porcelliDprime}
\eeq
These scalings are correct provided $\Delta'\din\ll1$, 
or $\Delta'\rho_s^{1/3}d_e^{2/3}\ll1$. At larger $\Delta'$, 
again, like in \apref{app:ultralowbeta}, 
$\din$ limits the magnitude of the current and we replace 
$\Delta'$ by $1/\din$ in \eqref{eq:lambdaalpha_gen}:
\beq
\lambda^2 \sim \alpha,
\label{eq:lambdaalpha}
\eeq
whence follow the scalings\cite{porcelli:425,zakharov:3285,mirnov:4468}
\beq
\delta\sim\frac{d_e^{4/3}}{\rho_s^{1/3}},\quad
\din\sim d_e^{2/3}\rho_s^{1/3},\quad
\gamma\sim k_y v_A\,\frac{d_e^{1/3}\rho_s^{2/3}}{L_s}.
\label{eq:porcelli}
\eeq

Checking now the condition $\alpha\gg1$, we find that the finite-$\Delta'$ 
scalings \exref{eq:porcelliDprime} are valid provided
\beq
\frac{\rho_s^2}{d_e^2} \sim \beta_e\,\frac{m_i}{m_e} \gg (d_e\Delta')^2 
\eeq
and the ``infinite-$\Delta'$'' ones \exref{eq:porcelli} hold for
\beq
\frac{\rho_s^2}{d_e^2} \sim \beta_e\,\frac{m_i}{m_e} \gg 1.
\eeq

\subsection{Asymptotic solution for the two-fluid tearing mode\label{app:twofluid_sln}}

In Refs.~\onlinecite{pegoraro:364,porcelli:425}, 
the scalings \exref{eq:porcelli} were derived in a more mathematical way
via a rather involved double-layer matching procedure in wavenumber 
space under the assumption of isothermal electrons
($\adex=1$). There are several other derivations (notably 
Refs.~\onlinecite{drake:2509,zakharov:3285}), also 
analytically quite cumbersome. 
Here we give a calculation that recovers the essential 
result without assuming isothermal electrons and 
at the minimal analytical cost. 

We start by noting that, using \eqsand{eq:Apar_chi}{eq:match}, we may write $\Apar$ 
in a form that is automatically matched to the outer solution:
\beq
\Apar = -\chi_0\lt[1 + \tchi + \frac{1}{2}\,\Delta'\din\xi + 
\xi\int_\xi^\infty dz\,\frac{\tchi'(z)}{z}\rt].
\label{eq:Apar_ion}
\eeq 
Our strategy will be first to solve \eqref{eq:chi_eq} in the ion 
region and use the solution $\tchi_i$ in the above formula for $\Apar$;
then to solve in the electron region and use the solution $\tchi_e$ 
in \eqref{eq:Apar_chi}; finally to take the large-argument asymptotic 
of the electron solution $\Apare$ and match it with the small-argument 
asymptotic of the ion solution $\Apari$. 

\subsubsection{Ion region}

The ion region is $x\sim\rho_s$, or $\xi\sim\alpha\gg1$. 
In this limit, the $1/\xi^2$ and $\lambda^2\tchi$ terms in \eqref{eq:chi_eq}
are negligible (the latter because we are anticipating 
the ordering $\lambda^2\sim\alpha$) and $\adex\approx1$. 
Introducing the new variable $z=\xi/\alpha\sqrt{1+\tau/Z}=x/\rho_\tau$, 
where $\rho_\tau=\rho_s\sqrt{1+\tau/Z}$, we can now write the 
rest of \eqref{eq:chi_eq} as follows
\beq
\tchi_i'' - \tchi_i = \frac{\tlambda^2}{z^2},
\label{eq:chi_ion}
\eeq
where we have denoted
\beq
\tlambda^2 = \frac{\lambda^2}{\alpha^2(1+\tau/Z)} 
\ll1. 
\eeq
\Eqref{eq:chi_ion} can be solved exactly, subject to the 
boundary condition $\tchi_i(z\to\infty)\to0$: 
\beq
\tchi_i = e^{-z}\lt[C_i - \tlambda^2\int_{z_0}^z dr\,e^{2r}
\int_r^\infty ds\,\frac{e^{-s}}{s^2}\rt],
\label{eq:chi_ion_sln}
\eeq
where $C_i$ is a constant of integration and the parameter $z_0$ 
in the particular integral can be chosen at will, with the difference 
absorbed into the homogeneous part of the solution (the constant $C_i$). 

Substituting the solution~\exref{eq:chi_ion_sln} into \eqref{eq:Apar_ion}, 
taking the limit $z\to 0$ and keeping only the leading-order 
contributions (in $1/\alpha$) in all terms of the expansion, 
we~get
\bea
\nonumber
\Apari&\approx& -\chi_0\biggl[1 + C_i 
+ \frac{1}{2}\,\Delta'\din\xi\biggr.\\
&&-\biggl. \tlambda^2\ln\xi + \frac{C_i}{\alpha\sqrt{1+\tau/Z}}\,\xi\ln\xi\biggr],
\label{eq:Apari}
\eea
where we formally ordered $\Delta'\din\sim1$ and anticipated $C_i\sim1$
(all of this will be checked {\em a posteriori}). 

\subsubsection{Electron region}

The electron region is $x\sim\delta$, or $\xi\sim1/\alpha\ll1$. 
In this limit, the $\xi^2\tchi$ term in \eqref{eq:chi_eq} is negligible 
and we are left with a homogeneous equation for $\tchi+1 = \chi/\chi_0$. 
Introducing the new variable $y=\sqrt{2}\,\alpha\xi=x/\delta$, we get
\beq
y^2\,\frac{d}{dy}\lt[\frac{1}{y^2} + \frac{1}{2}\lt(G(y) + \frac{\tau}{Z}\rt)\rt]\chi_e'
= \frac{\lambda^2}{2\alpha^2}\,\chi_e.
\label{eq:chi_el}
\eeq
The right-hand side is, in fact, small, and so we can solve this 
equation perturbatively, in powers of $\lambda^2/\alpha^2\sim1/\alpha$
(cf.~Ref.~\onlinecite{zakharov:3285}). 
We will see that we need to do this to second order: 
$\chi_e = \chi_e^{(0)} + \chi_e^{(1)} + \chi_e^{(2)}$. 
Our boundary condition will be that the current, proportional 
to $\chi'/y$, must be even as $y\to0$. We can also let 
$\chi_e^{(1)}(0) = \chi_e^{(2)}(0) = 0$ without loss of generality, 
so $\chi_e(0)=\chi_e^{(0)}(0)$. 

Integrating \eqref{eq:chi_el} to zeroth order (right-hand side~$=0$), 
we get
\beq
\frac{\chi_e^{(0)}}{\chi_0} = C_e,
\eeq
where $C_e$ is a constant of integration. 
To first order, we obtain, again by direct integration, 
\beq
\frac{\chi_e^{(1)}}{\chi_0} = -\frac{\lambda^2}{2\alpha^2}\,C_e 
\int_0^y\frac{du\,u}{1+ (u^2/2)\lt[G(u) + \tau/Z\rt]}.
\eeq
Finally, to second order, 
\begin{align}
\nonumber
\frac{\chi_e^{(2)}}{\chi_0} = -&\frac{\lambda^4}{4\alpha^4}\,C_e 
\int_0^y\frac{du\,u^2}{1+ (u^2/2)\lt[G(u) + \tau/Z\rt]}\\
&\times\int_0^u\frac{dv}{v^2}
\int_0^v\frac{dw\,w}{1+ (w^2/2)\lt[G(w) + \tau/Z\rt]}.
\end{align}

Substituting the solution worked out above into \eqref{eq:Apar_chi}, taking 
the limit $y\to\infty$ and again throwing out all terms 
subdominant in $1/\alpha$, we~find
\beq
\Apare \approx -\chi_0 C_e \biggl[1 + \IG \Lambda \xi 
- \tlambda^2\ln\xi + \IG\Lambda\tlambda^2\xi\ln\xi\biggr],
\label{eq:Apare}
\eeq
where $\Lambda = \lambda^2/\alpha\sqrt{1+\tau/Z}$ and 
\beq
\IG = \int_0^\infty \frac{dy\sqrt{1+\tau/Z}/\sqrt{2}}{1+(y^2/2)\lt[G(y)+\tau/Z\rt]}
\eeq
is a number of order unity. Had electrons been 
isothermal ($G=1$), it would have been $\IG=\pi/2$. 

\subsubsection{The dispersion relation}

We now match the asymptotics \exref{eq:Apare} and \exref{eq:Apari} 
term by term. The $\ln\xi$ term can be ignored as long as 
we carry out the matching for $\xi\gg1$ --- we can do this because 
the electron solution is in fact valid beyond the electron 
region ($\xi\sim1/\alpha$) all the way through the intermediate 
region $\xi\sim1$, or $x\sim\din$ (because we have solved 
to second order in $1/\alpha$). 

Matching the remaining three terms 
allows us to determine the two constants of integration and find the dispersion 
relation:\cite{pegoraro:364,porcelli:425} 
\beq
\frac{1}{2}\,\Delta'\din = \frac{\IG\Lambda}{1-\IG\Lambda^2}. 
\label{eq:disprel}
\eeq
This indeed has the two limits that we intuited in \apref{app:twofluid}: 
\begin{align}
\Delta'\din\ll1~&\Rightarrow~
\Lambda \equiv \frac{\lambda^2}{\alpha\sqrt{1+\tau/Z}} 
= \frac{1}{2\IG}\,\Delta'\din,\\
\Delta'\din\gg1~&\Rightarrow~
\Lambda \equiv \frac{\lambda^2}{\alpha\sqrt{1+\tau/Z}} 
= \frac{1}{\sqrt{\IG}},
\end{align}
so we now know not just the scalings but also the numerical prefactors
(which depend on the functional form of the effective adiabatic exponent 
$G$ of the electrons via the constant $\IG$). 
\Eqref{eq:disprel} gives a smooth connection between the two limits.

\subsection{Semicollisional tearing mode\label{app:tearing_coll}}

\subsubsection{Equations}

To treat the tearing mode in the semicollisional limit,\cite{drake:1341} 
we must use \eqsref{eq:phi_lin}, \exref{eq:Apar_lin} 
and \exref{eq:dTe_Sp_lin}, again taking $\kpar(x)\approx k_yf(x)$. 
The outer region is still MHD as described in \apref{app:outer}. 
In the inner region, instead of \eqsand{eq:phi_inner}{eq:Apar_inner}, we get
(the first equation is unchanged)
\begin{align}
\label{eq:phi_inner_coll}
&\frac{\delta}{x}\frac{Z}{\tau}(1-\hat{\Gamma}_0)\tphi =
\frac{1}{2}\,d_e^2\dd_x^2\Apar,\\
\label{eq:Apar_inner_coll}
&\frac{\delta}{x}\lt(1-\frac{\eta}{\gamma}\,\dd_x^2\rt)\Apar =
\lt[1 + \adex\,\frac{Z}{\tau}(1-\hat{\Gamma}_0)\rt]\tphi,\\
&\adex\lt(\frac{x}{\delta}\rt) = 
\frac{3 + (\kape\gamma/\vthe^2)(x/\delta)^2}{1 + (\kape\gamma/\vthe^2)(x/\delta)^2}.
\label{eq:adex_coll}
\end{align}

These equations turn out to have a mathematical structure that can be 
exactly mapped onto the collisionless case.\cite{cowley:3230} 
Let us introduce a new scale
\beq
\dres = \frac{\delta}{d_e}\lt(\frac{\eta}{\gamma}\rt)^{1/2}\!\!\! 
= \frac{L_s^2}{\sqrt{2}\,\rho_s}\lt(\frac{\gamma}{k_y v_A}\rt)^{1/2}
\!\!\!(k_y L_s S)^{-1/2},
\eeq
where $S=v_AL_s/\eta$ is the Lundquist number. Then, denoting 
$\Phi=(\dres/\delta)\tphi$, we can recast 
\eqsdash{eq:phi_inner_coll}{eq:adex_coll} 
in the following form
\begin{align}
\label{eq:phi_Delta}
&\frac{\dres}{x}\frac{Z}{\tau}(1-\hat{\Gamma}_0)\Phi =
\frac{1}{2}\frac{\eta}{\gamma}\,\dd_x^2\Apar,\\
\label{eq:Apar_Delta}
&\frac{\dres}{x}\lt(1-\frac{\eta}{\gamma}\,\dd_x^2\rt)\Apar =
\lt[1 + \adex\,\frac{Z}{\tau}(1-\hat{\Gamma}_0)\rt]\Phi,\\
&\adex\lt(\frac{x}{\dres}\rt) = 
\frac{3 + (a/2)(x/\dres)^2}{1 + (a/2)(x/\dres)^2},
\label{eq:adex_Delta}
\end{align}
where $a=2\eta\kape/d_e^2\vthe^2$. For the collision operator we 
chose in \secref{sec:colls}, $\eta = \nu_{ei} d_e^2$, $\kape=\vthe^2/2\nu_{ei}$, 
and so $a=1$; the isothermal closure would formally correspond to $a=\infty$. 
Comparing \eqsdash{eq:phi_Delta}{eq:adex_Delta} with \eqsdash{eq:phi_inner}{eq:Gvsx}, we see 
that all results obtained for the collisionless case can be converted 
into analogous results for the semiciollisional case by mapping
\beq
\delta\to\dres,\quad
d_e^2\to\frac{\eta}{\gamma} 
\label{eq:map}
\eeq
and using \eqref{eq:adex_Delta} instead of \eqref{eq:Gvsx} for the 
effective adiabatic exponent of the electrons. 

Again, these equations can be manipulated into a form derived in 
Ref.~\onlinecite{cowley:3230}. 
Proceeding analogously to the way we did in \apref{app:inner}, we get 
\bea
\label{eq:Cowley_Epar_coll}
\frac{x}{\dres}\lt(\Apar - \frac{x}{\dres}\,\Phi\rt)
\sigma\lt(\frac{x}{\dres}\rt) &=& \frac{Z}{\tau}(1-\hat{\Gamma}_0)\Phi,\\
\frac{1}{2}\frac{x}{\dres}\lt(\frac{\dres^2}{\delta^2}\,d_e^2\rt)
\dd_x^2\Apar &=& \frac{Z}{\tau}(1-\hat{\Gamma}_0)\Phi,\qquad\quad
\label{eq:Cowley_jpar_coll}
\eea
where the ``scaled effective conductivity'' is 
\beq
\sigma\lt(\frac{x}{\dres}\rt) = 
\frac{1 + (a/2)(x/\dres)^2}{2 + (3+a)(x/\dres)^2 + (a/2)(x/\dres)^4}.
\eeq 
\Eqsand{eq:Cowley_Epar_coll}{eq:Cowley_jpar_coll} are 
Eqs.~(76) and (78), respectively, of Ref.~\onlinecite{cowley:3230}, 
again in the special case of no equilibrium density gradient
(to recover their equations exactly, replace $\delta\to-i\delta$, 
$\dres\to\dres/\sqrt{2}$ and $\sigma\to\sigma/2$). 

Using the mapping \exref{eq:map}, we conclude that 
the general tearing mode equations \exref{eq:chi_eq} and \exref{eq:match}, 
derived in \apref{app:math}, now hold with 
\begin{align}
\label{eq:din_coll}
&\din = \lt(\sqrt{2}\,\rho_s\dres\rt)^{1/2} 
=L_s\lt(\frac{\gamma}{k_y v_A}\rt)^{1/4}\!\!\!(k_yL_s S)^{-1/4},\\
\label{eq:lambda_coll}
&\lambda^2 = \frac{\sqrt{2}\,\rho_s\dres}{\eta/\gamma} 
= \lt(\frac{\gamma}{k_y v_A}\rt)^{3/2}\!\!\!(k_yL_s S)^{1/2},\\
\label{eq:alpha_coll}
&\alpha = \frac{\din}{\sqrt{2}\,\dres} = \frac{\rho_s}{\sqrt{2}\,\dres} 
= \frac{\rho_s}{L_s}\lt(\frac{\gamma}{k_y v_A}\rt)^{-1/4}\!\!\!(k_yL_s S)^{1/4},
\end{align}
and $\adex(\sqrt{2}\,\alpha\xi)$ given by \eqref{eq:adex_Delta}. 
All the same mathematical considerations apply, with the 
(qualitatively inconsequential) 
exception that the functional form of the effective adiabatic exponent 
is different. 

We conclude by summarizing 
the heuristically obtainable scalings for the semicollisional case
--- these can be read off from the results of \apsand{app:tearing}{app:twofluid_sln}
with the aid of the definitions \exsdash{eq:din_coll}{eq:alpha_coll}. 
From this point on, all lengths are normalized by $L_s$. 

\subsubsection{Resistive MHD (one-fluid) limit}

Proceeding analogously to the case considered in \apref{app:ultralowbeta}, 
we assume $\alpha\ll1$, or $\rho_s\ll\dres$, and recover the classic 
resistive-MHD tearing mode.\cite{FKR,coppi-res,ABC} First, at finite $\Delta'$, 
\eqref{eq:lambda_gen} with the definitions \exref{eq:din_coll} 
and \exref{eq:lambda_coll} gives the scalings\cite{FKR} 
\bea
\dres &\sim& \Delta'^{2/5}\rho_s^{-1}(k_y S)^{-4/5},\\
\din &\sim& \Delta'^{1/5}(k_y S)^{-2/5},\\
\frac{\gamma}{k_y v_A} &\sim& \Delta'^{4/5}(k_y S)^{-3/5}.
\eea
The reconnection layer's width is $\din\ll\dres$ (because $\alpha\ll1$) 
and the condition for the above 
scalings to apply is $\Delta'\din \ll 1$, which translates into 
\beq
\Delta' \ll (k_y S)^{1/3}. 
\label{eq:finDprime}
\eeq
When this is broken, one gets the "infinite-$\Delta'$" scaling \exref{eq:lambda1}, 
whence follow the scalings\cite{coppi-res,drake:1777}
\bea
\dres &\sim& \rho_s^{-1}(k_y S)^{-2/3},\\
\din &\sim& (k_y S)^{-1/3},\\
\frac{\gamma}{k_y v_A} &\sim& (k_y S)^{-1/3}. 
\eea

The resistive MHD results are valid provided $\alpha\ll1$, which imposes an upper 
bound on the ion scale: 
\beq
\rho_s \ll \Delta'^{1/5}(k_y S)^{-2/5} 
\eeq
for the finite-$\Delta'$ scalings and 
\beq
\rho_s \ll (k_y S)^{-1/3} 
\label{eq:onefluid_coll}
\eeq
in the ``infinite-$\Delta'$'' limit. 

\subsubsection{Two-fluid case}

When the ion scale is sufficiently large ($\alpha\gg1$), 
two-fluid effects become important,
similarly to the case considered in \apref{app:twofluid}. 
Using \eqref{eq:lambdaalpha_gen} 
and the definitions \exsdash{eq:din_coll}{eq:alpha_coll}, 
one gets the scalings\cite{drake:1341,pegoraro:647} 
\bea
\dres &\sim& \Delta'^{1/3}\rho_s^{-2/3}(k_y S)^{-2/3},\\
\din &\sim& (\Delta'\rho_s)^{1/6}(k_y S)^{-1/3},\\
\frac{\gamma}{k_y v_A} &\sim& (\Delta'\rho_s)^{2/3}(k_y S)^{-1/3}.
\eea
These scalings hold provided $\Delta'\din\ll 1$, or 
\beq
\Delta' \ll \rho_s^{-1/7}(k_y S)^{2/7}.
\eeq
At larger $\Delta'$, \eqref{eq:lambdaalpha} must be used, 
whence follow the scalings\cite{drake:1777,pegoraro:364,aydemir:3025} 
\bea
\dres &\sim& \rho_s^{-5/7}(k_y S)^{-4/7},\\
\din &\sim& \rho_s^{1/7}(k_y S)^{-2/7},\\
\frac{\gamma}{k_y v_A} &\sim& \rho_s^{4/7}(k_y S)^{-1/7}.
\eea

In both cases, the width of the reconnection layer is $\dres\ll\din$ 
(because $\alpha\gg1$), which holds if 
\beq
\rho_s \gg \Delta'^{1/5}(k_y S)^{-2/5} 
\eeq
for the finite-$\Delta'$ scalings and 
\beq
\rho_s \gg (k_y S)^{-1/3},
\eeq 
in the ``infinite-$\Delta'$'' limit. 


\end{document}